\pdfoutput=1
\documentclass[useAMS,usenatbib]{mn2e}
\usepackage{deluxetable}
\usepackage[pdftex]{graphicx}
\usepackage[pdftex,usenames,dvipsnames]{xcolor}
\usepackage{amssymb}
\usepackage{aas_macros}
\usepackage{url}
\usepackage[none]{hyphenat}
\usepackage{times}
\usepackage[linktocpage=true,backref=true,pagebackref=false,bookmarks=true,bookmarksnumbered=False,colorlinks=true,linkcolor=NavyBlue,citecolor=NavyBlue,urlcolor=NavyBlue]{hyperref}
\pdfpageattr {/Group << /S /Transparency /I true /CS /DeviceRGB>>}  

\usepackage[backgroundcolor=yellow,linecolor=red,bordercolor=yellow,textsize=scriptsize]{todonotes}
\pdfoutput=1
\title[From kpcs to the central parsec]{From kpcs to the central parsec of NGC 1097: feeding star formation and a black hole  at the same time\thanks{Based on observations from VLT 60.A-9026A, 77.B-0728A, HST13413, ALMA 2011.0.00108.5}}

\author[M.A. Prieto et al.]{M. Almudena Prieto$^{1,2}$\thanks{Email: aprieto@iac.es}, Juan A. Fern\'andez-Ontiveros$^{1,2}$, Gustavo Bruzual$^3$,\newauthor
 Andreas Burkert$^{4,5}$, Marc Schartmann$^{4,5}$, Stephan Charlot$^{6}$
 \\
$^1$Instituto de Astrof\'isica de Canarias (IAC), E--38200 La Laguna, Tenerife, Spain.\\
$^2$Universidad de La Laguna, Dept. Astrof\'isica, E--38206 La Laguna, Tenerife, Spain\\
$^3$Instituto de Radioastronom\'ia y Astrof\'isica, UNAM,  C.P. 58089 Morelia, M\'exico\\
$^4$Universit\"ats-Sternwarte M\"unchen, Scheinerstr. 1, D-81679 M\"unchen, Germany \\
$^5$Max-Planck-Institut f\"ur extraterrestrische Physik, Postfach 1312, Giessenbachstr., D--85741 Garching, Germany\\
$^6$Sorbonne Universit\'es, UPMC-CNRS, UMR7095, Institut d'Astrophysique de Paris, F-75014 Paris, France\\}
\date{}

\pagerange{\pageref{firstpage}--\pageref{lastpage}} \pubyear{2018}

\def\LaTeX{L\kern-.36em\raise.3ex\hbox{a}\kern-.15em
 T\kern-.1667em\lower.7ex\hbox{E}\kern-.125emX}

\begin{document}
\label{firstpage}
\maketitle

\begin{abstract}
A panchromatic view of the star forming ring and  feeding process  in the central kpc of 
the galaxy NGC 1097 is presented. The assembled IR to  UV images at $\sim$10 pc resolution allow us to characterise the population of circa 250 clusters in the ring and  disentangle the network of  filaments of dust and gas  that enshroud and feed them. 
The ring is a place of intermittent star bursts over the last 100 Myr. Four major episodes covering a proto-cluster phase of eleven mid-IR sources  at  the molecular clouds core, and two (three) previous bursts  with a time separation of   20 - 30 Myr are identified.
     The extinction map of the inner few kpc  resolves NGC1097's  two major dust lanes  in bundles of narrow, $<$25 pc width,  filaments  running  along  the galaxy's bar.   As they approach the ring, some    circularise along it,  others curve  to the centre to produce a nuclear spiral. We believe these are   kpc-scale dust-gas streamers  feeding the ring and the black-hole.
The total mass in clusters formed  in the ring  in the last 100 Myr is  $< 10^7\, \rm{M_\odot}$,  i.e.  $< 1\% $ of the  $10^{9} M_\odot$ of molecular gas in the ring; yet, at its current star formation rate,  $\sim1.8\, \rm{M_\odot \, yr^{-1}}$, an order of magnitude more in stellar mass should have been produced over that period. This means that  the availability of gas  in the ring is not  the sole star formation driver, perhaps the rate at which dense   gas accumulates in the ring   is key.


\end{abstract}


\section{Introduction} \label{sec:intro}

Feeding and feedback in galaxies affects  two major phenomena in their centre: the growth of super massive black holes (BH) and their activity cycles,   and the  on-set of new star formation. The supplying material to sustain those processes, the mechanism by which   material falls into the centre are standing issues. The environment of galaxies whether in clusters or in the field, and their history via mergers and interactions,  are phenomena that will naturally lead to the rain of fresh material into the centre providing a substantial reservoir for new star formation and BH feeding (e.g. Silk \& Rees 1998; Gebhardt et al. 2000; Springel et al. 2005; Fabian 2012)

High angular resolution observations particularly at frequencies not subjected to dust extinction are  providing detailed  insight  on the feeding process  in the centre of galaxies at scales of  few parsec, which mainly occurs via narrow dust / gas streamers (e.g. Mueller-Sanchez et al. 2009, Mezcua et al. 2015, Imanishi et al. 2016 for NGC 1068;   Espada et al. 2017 for Cen A; Mezcua et al. 2016 for Circinus;  
Storchi-Bergmann et al. 2010 for NGC 4151; Malkan et al.1998, Prieto et al. 2014,    for samples of the nearest Seyfert galaxies), via nuclear spirals (e.g. Prieto et al. 2005, Davies et al. 2009, Fathi et al. 2013 for NGC 1097; Combes et al. 2014 for NGC 1566; Combes et al. 2018),   via  G2-like clouds $ "a~la ~SgA* "$ (Gillessen et al. 2012 at the Galactic Center).  On galactic  scales, bars, mergers  are among  the most  efficient mechanisms   to drive material to the central kpc where  it often ends up in a circumnuclear star forming ring. The latter is  supported  by observations of barred galaxies which often present large molecular gas concentrations  and  ensued star formation  (e.g. Matsuda \& Nelson 1977;  Sakamoto et al. 1999).
Still, the  comprehensive  view of the complete  feeding process from the source of the feeding material  to where and how  ultimately it is deposited   relies  mostly on theoretical predictions and simulations (e.g. Shlosman et al. 1990;  Friedli \& Benz 1993; Springel et al. 2005; Li et al. 2015 and references therein).

 This work relies   on multiwavelength - UV to mm-  high angular resolution data  to illustrate  the feeding process and star formation history of the very bright and highly populated circumnuclear star forming ring  in the barred galaxy NGC 1097. The  stellar ring in this galaxy is a reference because of the large number of clusters, in the few hundreds, it contains and their high brightness. Yet, the galaxy is sufficiently close to individually resolve 
the clusters at scales of parsecs. 

NGC1097   is  an early-type , barred,  spiral with   two grand-design spiral arms. Each arm    
    anchors  at the North and South edges of the galaxy disc,  at which  point  about 10 kpc  long dust-lanes develop and  extend  straight up to the central kpc of the galaxy  along  a bar seen in HI.  The Northern  spiral arm is  distorted at its end at the location of  the companion galaxy NGC 1097A. The total mass of the galaxy in HI is  $M_{\rm HI_{total}} = 5 \times 10^9\, \rm{M_\odot}$ (Ondrechen et al. 1989).
    
At the central $\sim$ 800 pc  radius, a very bright circumnuclear  star forming ring stands out. The ring  resolves in hundreds of  stellar clusters seen   prominent in the  optical  (e.g. Barth et al. 1995, this work), near and mid-IR (e.g. Prieto et al. 2005, Mason et al. 2007; Reunanen et al. 2010), cm (Beck et al. 2005),  X-rays  and  in molecular  $H_2$ (Mezcua et al. 2015), CO and HCN lines among others (e.g. Izumi et al. 2013, Martin et al. 2015), barely in  HI (Ondrechen et al. 1989).

 The nucleus hosts a BH mass  of $M_{\rm BH} = 1.2 \times 10^8\, \rm{M_\odot}$ (Lewis \& Eracleous 2006),  with    moderate activity given the low  bolometric luminosity of the source extracted from integration of the central 10 pc spectral energy distribution (SED):     $L_{\rm bol} \sim 4 \times 10^{41} erg~s^{-1}$ (Prieto et al. 2010), which places it as a very   low efficiency source accreting at a rate of $ < 3\times 10^{-5} $ 
in Eddington units.  The  nucleus has shown activity cycles, the optical spectrum mutating  from a low ionisation --\,LINER type\,--    to a double-peaked broad line, type 1 source   (Storchi-Bergman et al. 1997). 
 
 The nature of the nucleus emission as inferred from the analysis  of  the central 10 parsec-scale spectral energy distribution (SED) is   synchrotron jet-dominated  emission    (Fernandez-Ontiveros et al. 2012 and in preparation, Koljonen et al. 2015). The core is  developing a pc-scale jet discovered   with VLBA in 8.4 GHz (Mezcua \& Prieto, 2014). The radio loudness of the core, measured as F(5 GHz) / F( 2180 A ) from our 10 pc scale SED (see Mezcua \& Prieto, 2014) is about 200, in line with what  often found in low efficient accreting sources (Sikora et al. 2017.
 

This paper makes use of  subarcsec UV  to IR  data  to spatially resolve  the ring  in its individual components: the star clusters and  the network of dust / gas filaments in which they reside. 
The panchromatic view at a similar scale of $\sim 10 pc$ across the electromagnetic spectrum allows us  1) to resolve   spatial and temporally the star formation history in the ring via the individual analysis of the  cluster population; 2) to spatially resolve the network of narrow dust filaments that enshroud the ring. The  nucleus of NGC 1097 is  known to  be enshrouded by  also a  network of narrow dust-filaments that spiral   from the ring all the way to  the galactic centre  (Prieto et al. 2005). This work connects the nuclear dust spiral with  the filaments in the ring and further out  at several kpc distance with the two major dust lanes of the galaxy where they all origin. Characteristic physical parameters of the filaments are derived. Putting together all the information, a comprehensive view  of the origin, feeding and recurrent star formation in the central kpc of the galaxy is attempted.

%
A distance to NGC 1097 of 14.5 Mpc (Tully 1988) is used,    1 arcsec = 70 pc.



\section{Multiwavelength dataset }

The data comprises  of subarcsec resolution continuum- and hydrogen-recombination-lines- images spanning  the $0.3$--$20\, \rm{\micron}$ range, complemented with arcsec-scale submillimeter data at 88 GHz (Fig. 1). The sources are    \textit{Hubble Space Telescope} (\textit{HST}) images from   Wide-Field Camera 3 (WFC3)  in U-, B-, V-, I- , H$\alpha+[\textsc{N\,ii}]$ and adjacent continuum- bands, namely in the filters  F330W, F438W, F547M, F814W, F657N and FR656N respectively, and  from NICMOS  in Pa$\alpha$ and adjacent-continuum,  F187N and F190N filters respectively. The final multi-drizzle, cosmic-ray-free images delivered by the HST pipeline  are used in this work. 

The dataset further includes ESO-Very Large Telescope (VLT) data from the Adaptive-Optics-assisted  NACO camera in J-, H- , Ks- and L-bands from Prieto et al. (2005) and  diffraction-limited, line-free VISIR  images  in narrow-band  continuum filters centred at $11.88\, \rm{\micron}$ and $18.72\, \rm{\micron}$ from Reunanen et al. (2010).

 The NACO images have angular resolutions FWHM $\lesssim$ 0.15 arcsec (Prieto et al. 2005) comparable to those of the HST in the optical; the VISIR images have 
 FWHM $\lesssim$ 0.35 arcsec (Reunanen et al. 2010).
HST and VLT images  were all  registered to a common reference system using  nine, bright and isolated   clusters in the stellar ring, present in all the  images. The achieved registration precision  is $\sim$ 30 mas (see Mezcua et al. 2015 for details).

The set is further complemented with archived ALMA data in band 3 (centred at 88 GHz), from which  the   HCN(1-0)  88.6 GHz line map from the calibrated data cube in the archive
 was extracted. This line was selected for being a high  density tracer and among  the strongest molecular line in the ring  (Martin et al. 2015). The HCN line-map has  a  beam resolution of $\sim 2.2 \times 1.5\, \rm{arcsec^2}$,   a factor three to ten worse  than that of the UV - IR dataset. Accordingly,  the registration of this image   with  the UV - IR dataset had to rely  on  NGC 1097's nucleus. Nonetheless, the registration precision  is   high as the eleven  point-like sources detected in the VLT-VISIR$18\, \rm{\micron}$ image fall at  corresponding  emission peaks in  the HCN map (Fig. 1).

 To complete the panchromatic view,  the deepest available  X-rays Chandra image of  the central kpc of the galaxy  is shown on top of the HST-UV image in Fig. 1. The nucleus is the strongest point source in the field (discussed in Mezcua et al. 2015) and the second clear detection  is a point-like source at $\sim$  4 arcsec South-West from  the nucleus  with no cluster counterpart.  The ring itself shows  some low level of diffuse emission across, somewhat pinpointing a few of the brightest regions in the ring, but  the signal to noise level is very low which unfortunately  warrants any  further analysis. We are hoping getting much deeper X-ray image, thus no further discussion of this data is in this work. 


\section{Star clusters  identification}

On the registered UV - IR dataset, a total of 247 individually resolved star clusters detected above a 3 sigma level with respect to their local background were identified. The reference  identification was performed at the ESO/NACO Ks-band to overcome dust extinction, and separately also in the HST / F336W U-band  to enhance cluster contrast over the  background bulge light that dominates in the IR. Cross identification with the rest of the images followed next.
Of the 247 sources, we secured 171 clusters detected in the UV (HST U-band), and up to $ 2\, \rm{\micron}$ (VLT NACO-Ks-band) most of them,  with a few  also  detected up to $4\, \rm{\micron}$ (VLT NACO L-band, detection  is limited because of  sensitivity issues). 
The  remaining 76 sources are detected from  $0.8 \, \rm{\micron}$  onward  only  because of  dust extinction. 

There are  a few detections of clusters outside the strict ring limits, e.g. in  between the nucleus  and the ring, or beyond  1 kpc radius. They all show properties similar to the rest of the sample in terms of size, luminosity,  mass, age,  and for the purposes of this work they will be treated as  integral part of the total  sample. 

Fig. \ref{ha}   shows a contours-version of the  HST-UV image   after removal of the   stellar light  using  a filter (white - tophat) that remove the low frequency diffuse light thus enhancing  point-like sources in the image. Accordingly, Fig. \ref{ha}  singles out  the cluster population  identified in the UV only (note that  the UV- and K- band images  in Fig. 1,  show  the complete cluster set detected in both UV  and /or in  K- band). Removing  of the stellar light enhances  the clusters contrast thus facilitating their identification and isolation with  respect to neighbouring ones.  Once  clusters are identified,  photometry  is done  on the original  image without any filtering applied.


\subsection{Cluster identification in HII gas } 

Cluster identification was equally pursued in the HST WFC3-H$\alpha$+[NII]- and HST NICMOS-Pa$\alpha$ images after continuum emission subtraction (sect. 3.3, continuum-free line images are shown in Fig. \ref{ha}).
Prime identification was done in the H$\alpha$+[NII] image that have comparable resolution to that of the UV and near-IR set.  The nebular gas shows  diffuse emission all over the ring,  and few clusters were found to have a significant point like counterpart in  gas.  Most do not, this is a priory consistent with the ages of the large fraction of clusters in the ring, $ >10 ~Myr$  (discussed in sect. 4) but it is somewhat unexpected for the youngest   $\sim < 4 ~Myr$ old population, for which  the detected H$\alpha$ emission is lower than what is predicted for their age.
The deficit in H$\alpha$ emission for this population  is confirmed after further examination of a second although less deep HST  ACS-H$\alpha$+[NII] image,\footnote{Fathi et al. (2013) refer in their Fig.1 to an HST H$\alpha$ image that looks  different   from that  in Fig. \ref{ha}. The discrepancy is presumably due to the fact that  their image does not have the continuum emission subtracted,  thus all the  point-like sources  in their image are  predominantly   continuum rather than   H$\alpha$ emission. }  and of the HST NICMOS-Pa$\alpha$ image of the ring. All show  equivalent appearance.  It should be noted that the resolution of the NICMOS-Pa$\alpha$  image is three times lower than that of H$\alpha$ which hampers a detailed comparison. Still, the  Pa$\alpha$ image uncovers  a few regions   extinguished by dust in H$\alpha$, but nonetheless  the comparison of H$\alpha$  and  Pa$\alpha$ images in Fig. \ref{ha} shows consistent results.

%
%

\subsection{Clusters size}

The clusters's emission is unresolved   in all the images, thus cluster's sizes are   set by the achieved spatial resolution and are upper limits. These   are determined  as the Full-Width-Half-Maximum (FWHM) of a Gaussian function fit to the cluster light profile. The fit was applied to   the brightest, best isolated clusters in the ring. At the two extremes of the wavelength range,  the FWHM of point-like clusters in the WFC / F336W image is $0\farcs12 \lesssim 8 ~pc $; that in the  NACO~ / ~K-band is   $0\farcs14 \lesssim 12 ~ pc$. Thus, an upper limit to the clusters size is set to   $FWHM_{\rm clusters} < 8 ~pc$.  For comparison,  the size of the spatially resolved cluster emission in one of the nearest   starbursts, NGC 253, is FWHM$\sim ~$1.5 pc (Fernandez-Ontiveros et al. 2009). 

Cluster photometry in the 0.3 -- 5$\, \rm{\micron}$ region  was done  in  an aperture radius r = $0\farcs15$, consistent with the mean angular resolution of this spectral band, and after subtracting the local, median, background from an annulus with radii from $0\farcs2$ to $0\farcs3$.

\subsection{Gas extinction in the ring}
Gas extinction  was derived from the HST recombination map   H$\alpha$/Pa$\alpha$ (Fig. 1). H$\alpha$ was extracted from the WFC3 F657N image, with the continuum subtracted from a linear interpolation between WFC3 -F547M and -F814W images;  
Pa$\alpha$  from the NICMOS F187N image with the continuum subtracted from  NICMOS / F190N image. Prior  producing the ratio map, the angular resolution of the WFC3 H$\alpha$+[NII] image  was degraded to match that of NICMOS-Pa$\alpha$, and 
an estimate of  40\% [NII] 6548, 6584 \AA \ light contribution to the    H$\rm \alpha$+\rm[NII] blend (following Phillips et al. 1984) was subtracted from the  H$\alpha$+[NII] image. Witt et al. (1992) extinction curve and   recombination case B, $\rm Pa\alpha / \rm H\alpha = 0.116$ (Osterbrock 1989) applies.

The average  extinction in the ring is  $A_V \lesssim< 2\, \rm{mag}$,  in line with the high percentage of clusters,  $\sim 70\% $ of the total sample of identified clusters,  detected in the UV.  Exceptions  occur at locations next to the thicker dust filaments  in the ring and   at  the entrance of the   dust lane at the Southern region in  the ring  where  $A_V$ rises to  $6\, \rm{mag}$. Dust obscured regions in the ring  - filaments and  lanes - are easily visualised in the UV image as well as in the dust extinction  map discussed in sect. 5 (see also Fig.1), and 
in general, Av values in the gas are   at most regions  comparable with those applicable to the continuum light,    the latter being inferred from  the stellar population analysis   (sect. 4) or  from the continuum  extinction maps (sect. 5).

\section{Cluster properties: age, mass and extinction}
Dating of the cluster population  is  restricted on the  171 clusters that have secure identification  in the UV,  and thus cover the widest wavelength range  in this work, namely $0.3$--$2.5\, \rm{\micron}$ (up to $4\, \rm{\micron}$ in some cases). Spectral Energy Distributions (SED) were constructed for each of these UV clusters that have   detection in at least four, and up to nine, photometric bands. Most fits are based on 6 -7 photometric bands always covering the whole  UV to IR range.   Ages and masses were determined by SED fits using   Charlot \& Bruzual latest  update of their Single Stellar Population (SSP) models. 


The Charlot \& Bruzual's SSP models  whose astrophysical ingredients are in Gutkin et al. (2016), Wofford et al. (2016) are based on Kroupa's (2001)  Initial Mass Function (IMF). The results from these models were checked with the particular case in which  the IMF is populated stochastically following the prescriptions in Bruzual (2010).  The stochastic effect  was found relevant for clusters of mass  $< 10^4$ M$_\odot$, which is   not  the case for the bulk of the  population in the ring (see below). Thus the results discussed in this work follow   Charlot \& Bruzual's  models with fixed IMF.
%

The models used are based on   solar metallicity,  
and an age range   from 10$^4$ yr to 9.75 Gyr. The sampling in ages is progressively increasing with ages:    $1 - 5 \times 10^4 yr$  for ages less than $10^5 yr$,   0.1 Myr step  up to  10 Myr,   0.5 Myr step in the 10 -15 Myr range, 1  to 3 Myr step in the 15 -  60  Myr range, 10 Myr step from 60 Myr to few 100 Myr. The SSP fits are done  with  the whole range of ages covered  by the models rather than introducing an arbitrary cut in age. This choice is important taking into account   the wide range of ages covered by the ring. 

Because of the diffuse nature of the $H\alpha$ emission (sect. 3.1), fitting of the nebular emission  was not included in the SSP fits. However, for comparative purposes, 
per each model in  the Bruzual \& Charlot's  library,  H$\alpha$ nebular emission    associated with the ionising continuum budget    was calculated following   Osterbrock 's  (1989) prescription - this is relevant for the youngest ages, below 5 Myr. Bruzual \& Charlot's  models already include the stellar H$\alpha$ contribution in absorption, thus, the net difference between the absorption and emitting H$\alpha$ flux components  was ultimately compared  with the measured  H$\alpha$ at the cluster locations.     As the observable is via  a filter that includes   $H\alpha$ and continuum, prior to the comparison the  model derived H$\alpha$ flux was   convolved    with the HST-WFPC3 F657N filter used in the observations.


Cluster's SED  fitting was done  via a $\chi^2$ minimisation.  Free parameters in the fit are the cluster mass, age, and continuum dust extinction.  Examples of SSP fits are in Fig. \ref{fit}.

The simultaneous  inclusion of UV, optical and IR data  in the fit was found to be critical for the age determination in the particular range covered in the ring,   in the 1 Myr  to a few 100 Myr (see below). Moreover,  the moderate  extinction at the clusters location (sect. 5)  also helped in better constraining ages   as the observed  SED is close to its intrinsic shape  and minimum correction due  to dust is required in the fit. On this basis, special features in the SED  that were found  determinant for the age estimates are:  the steep UV-optical-IR  spectral slope  for assuring the ages of the younger, less than 10 Myr old population (in this case the cluster H$\alpha$ emission cannot be  used - sect. 3.1 - but the wide wavelength coverage  up to $2\, \rm{\micron}$ and minimum dust extinction makes the result very solid); the Balmer break  for ages older than  20 Myr;  the near-IR bands helped to  constrain the oldest  population   beyond   60 Myr because of   the  progressive  curvature of the SED, simultaneously,  towards  the  UV and IR bands.  

Dating of the  clusters    detected only  from $0.8\, \rm{\micron}$ onwards  was found  largely uncertain: the SSP fits are  based on four spectral bins only and the rather featureless  shape of the continuum  in the workable range, $0.8 - 2.5\, \rm{\micron}$,  is  alone insufficient to properly constrain the age. For this reason, these clusters are not further analysed in this work. 
 The denser dusty environment of these clusters is  evidenced  by comparing the UV- with the K-band image of the ring (Fig. 1), to see that  these clusters are often located   behind the  thicker dust filaments.

Putting in context   the cluster properties  of NGC 1097's ring  as  compared with    those of massive young star clusters in the Milky Way and nearby galaxies indicates very normal properties. 
 Cluster masses  are found  in the    
 $10^{4-5}\, \rm{M_\odot}$ range,   and thus the ring  classify as a population of Young Massive Star (YMS) clusters as those found in our Galaxy and nearby galaxies (Portegies Zwart et al. 2010).
Ages  span  in the    Myr to  a few 100 Myr range. For comparison, YMS in the Milky Way have ages in the  2 Myr to  20 Myr range,  in the LMC and Andromeda, up to a few 100 Myr  (Portegies Zwart et al. 2010 and references therein).
The range of continuum extinction found in the fit is  $ 0 < A_V < 2 \, \rm{mag}$, in line with the  continuum dust- and recombination- maps (Fig. 1, sects. 3 \& 5).



\subsection{Recurrent star formation  in the ring}
The  HST-F336W and VLT/NACO Ks-band images  in Fig. 1  show  the cluster population  colour coded according to the age. The optical and IR images in between these two  bands are similar   and not shown. 
The color code follows a continuous sequence as follows:  blue are the youngest population in the ring, less than 10 Myr old, green are those  between 10 and 40 Myr, in red those older  than 40 Myr and up to a few 100 Myr, in white, those detected longward of $ 0.8\, \rm{\micron}$ only because of extinction. These "white" clusters  are not dated (sect. 4). 

A histogram    in Fig \ref{bursts}a shows the relevant peaks in the distribution of ages.  Focusing on  continuous bins that sum up together   20  clusters or more,   about three main age periods are distinguished:  the most prominent is the burst   at $4 \pm1$ Myr, comprising $ \sim > 33\%$ of the  total   171 clusters  dated; a  second period centres in the   $\sim 20 - 40$  Myr range with $\sim18\%$ of the total, the uncertainty   is at least 10 Myr, the   step in the models;   a third period resumes, arguable, the  two peaks at  $\sim$ 60 Myr and  $\sim$ 90 Myr, comprising 23\% of the total.  The dispersion in ages in the  last two periods is large, chiefly caused by the rather similar shape of the models within these   ages.    

The histogram   shows also some minor peaks at extreme ages: below 1 Myr,   and  beyond 1 Gyr. There is also a small group at $\sim10~Myr$. All these clusters are  detected in the UV - optical range but not in the IR, the reason being that they are slightly fainter than the rest and their IR counterpart, particularly for the younger ones, drops steeply  with increasing wavelength  and gets    below our detection limit in the IR with  VLT-NACO. Judging from a comparative   analysis with the rest of clusters, we believe the clusters  below 1 Myr pertain to the 4 Myr group, and those in the Gyr range pertain to the 50 - 100 Myr group. Nonetheless, they  require further age diagnosis.

To asses the    range  of uncertainty in the distribution of the  main  burst periods identified,   Fig \ref{bursts}b shows    a highly censored age histogram in which   the assigned age to each cluster is the median of its 10 best fits  ranked by their  $\chi^2$.  It can be seen that this new histogram is a  smoothed version   of  the  one in Fig \ref{bursts}a: the age peak at 3 - 5 Myr still remains well defined and  isolated, but the two older periods merge into a smoothed distribution,  still bimodal, with   two main peaks at $\sim 30$ Myr and $\sim$ 50 Myr, the former with a dispersion of 10 Myr in line with the age-step of the models, the later presenting a broad tail  up to 100 Myr, both peaks  in any case remaining  within the  star formation  periods  identified in Fig \ref{bursts}a. 

Overall, a time separation of about 20 - 30 Myr between major burst periods is apparent. If so, the proto-cluster population may  still be  in its infancy (sect. 4.2). It may nonetheless   be possible that the  older burst periods include themselves a finer temporal sequence   of  bursts that   our  simple SED fit approach cannot   resolve.

Fig. \ref{fit} shows  examples of the SSP fits for representative cases of the various   age periods. Errors in the measured fluxes (circles) are smaller that the symbols and  not distinguishable,  upper limits apply for non detections. Age, mass and  Av derived from the fit are indicated in the plots.   It can be noted that the shape of the observed continuum changes  markedly between the different age periods identified: extremely blue and steep for the youngest,  less than 10 Myr old group,  an increasingly strong and broad IR SED and pronounced Balmer break in the 20 - 40 Myr case, a strong curvature in the SED in both  the UV and IR bands  for the oldest group. The predicted and measured H$\alpha$ emission is also shown. As discussed in sect. 3,  most of the clusters do not show  nebular emission, which is   consistent  with their assigned age but  for the youngest, less than 4 Myr old, population which should show relevant H$\alpha$ emission but do not (see sect. 3.1). Fig. \ref{fit} provides examples of this young population which is typically characterised by  very blue  SED,  low Av but   H$\alpha$ emission lower  than the prediction: two cases with lower emission than predicted, one case  in agreement with the prediction, are shown.

\subsection{The proto-cluster population}
Beyond the identified clusters  in the ring, an even younger  population  - presumably in a proto-cluster phase -  is identified  in the mid-IR. This population singles out  in VLT-VISIR diffraction-limited  images  of the ring at 11 and   $18\, \rm{\micron}$ (Reunanen et al. 2010).   Fig. 1d shows the VLT $18\, \rm{\micron}$ image, for best signal to noise ratio, to resolve  in circa ten - eleven, bright, point-like sources (dark-reddish point-sources in the image). The VISIR   $11\, \rm{\micron}$ show the same sources (not shown here but it is in  Reunanen et al. 2010) as well as  in an equivalent $11\, \rm{\micron}$ Gemini image published   in Mason et al. (2007).
In  Fig. 1d, it can be noticed that the location of these sources  just coincides with  the  emission peak   of an equivalent number of  ALMA / HCN(1-0) molecular clouds  resolved in the ring: almost each molecular cloud in the ring encloses one or two mid-IR sources. The sources also coincide  with the 
CO(1-0) clouds after comparing with  Hsieh's et al. (2011) map. Because of their strategic location,   these mid-IR sources are presumably   dust cocoons at the innermost -densest- region in the molecular cloud,    the factories where the new burst of star clusters is to emerge.

An estimate of their temperature  is inferred from fitting a grey black-body  to their $11$--$18\, \rm{\micron}$ SED. Assuming a standard dust emissivity with a power law exponent $\beta = 1.6$
( e.g. Barvainis 1987),  a grey black-body (BB) fit,  $\nu^{1.6} \times \rm{BB(T)}$,   yields a temperature, $T \lesssim 150\, \rm{K}$  for  most of  the sources. In a few cases it was possible to  associate the VISIR emission  with  ALMA sub-millimeter continuum emission  at $860 \, \rm{\micron}$ (this ALMA map is shown in Izumi et al. 2013), and  $3000 \, \rm{\micron}$ (this maps is not shown here but it is similar to the $860\, \rm{\micron}$ map), but the sub-millimeter emission turned to  be very steep, presumably it is dominated by the  strong  non-thermal component of the ring found in Tabatabaei et al. (2017),  and  could not be used to further constrain the dust  temperatures.

These mid-IR sources are furthermore found to be shielded from the  external radiation by highly optically thick material. The simple comparison between their associated  BB emission at  $2  \, \rm{\micron}$  with that  of the clusters   at this frequency implies  an  extinction  Av  $\sim 80$ mag. This is in line with the inferred column densities of the  molecular clouds,  $N_{H_{2}} \sim 10^{23} cm^{-2}$ (Hsieh et al. 2011),
which imply extinctions  of at least Av = 50 mag.

The size of these proto-clusters is  limited by the spatial resolution of our VISIR data, currently at the VLT diffraction limit at $11 \, \rm{\micron}$, to be $FWHM < 60~ pc$.


\subsection{On the diffuse nature of the clusters HII gas}

As illustrated in Fig. \ref{ha}, nebular  H$\alpha$ and Pa$\alpha$ are found all over the ring  but its structure is dominated by diffuse emission, no clear point-like counterpart  emission at the cluster locations, especially  for the youngest 4 Myr old population, is detected (sect. 3.1). Thus,    this population  shows a substantial deficit in HII gas as compared with  what is predicted for their age.

A further  characteristic of the HII gas   is its occasional filamentary morphology  at locations  away from the cluster positions,    for example at locations  North- and   South-West-  of the ring. Diffuse arch-like features sometimes extending 
$\sim$ 200 pc away from the ring (e.g. North of the ring) are apparent (Fig. \ref{ha}).

It thus follows that the HII  gas   measured at the cluster spatial scales  may not  be  indicative of the star formation  rate, nor of the age  of the cluster - an effect  also pointed out by  Hollyhead et al. (2015) who observe a similar phenomenon in the cluster population of  M83. For  the youngest 4 Myr old population in the ring of NGC 1097, the measured $H\alpha$ flux  would be inconsistent with  that age, however the   steep, extremely blue UV to IR spectra of this population could only be reproduced by SSP models of those ages. 
A stochastically populated IMF  could account for these spectra if     few OB stars are in place but stochastic effects  would explain sporadic cases and not what it seems to be a systematic.
 
Possibilities include  that the clusters are matter bounded which  would naturally lead to a leakage of photons, therefore a deficit in H$\alpha$ emission (Papaderos et al  2013). Still, one would expect  in this case a sharp  boundary in the  ionised gas at the location of the young clusters, which is  not the case as  no point-like counterpart in nebular gas are found. 

A possible explanation relates to a  low density cluster environment, low density allows for  the ionising photons to escape / distribute far beyond the cluster boundary. 
As a  test,  the size of the sphere of Stromgren was evaluated  at the clusters location. For HII region densities of $10^3 - 10^4$ $cm^{-3}$,  the size is   in the 2 - 3 pc range, i.e. smaller than the typical separation  between clusters in the ring,  in    the 10 pc range.  But if  gas densities are as low $100 - 10 ~cm^{-3}$ - as it appears to be the case in Giant Extragalactic HII regions  (Arsenault \& Roy 1988),   the Stromgren's sphere could be as large as 10 - 100 pc radius (Osterbrock, 1989), far beyond the stellar cluster boundary, with the effect that  overlapping spheres 
of  neighbouring young clusters  would produce the   diffuse appearance of the gas in the ring.


\section{The network of dust filaments: from the kpcs-scale dust lanes to  the central parsec} 

Lanes and filaments all over  the  ring are    visualised in the HST-UV image, and in its comparison with the almost dust-free  VLT-K-band image (Fig. 1a, b).  The  UV image also traces well   the known  nuclear spiral of  dust filaments, a sharper view of it in the VLT-NACO / J-band is shown for comparison purposes in Fig.1f taken from Prieto et al. (2005).

A   highly contrasted dust map of the central kpc of the galaxy   is provided via  the extinction map Av  in Fig. 1e. The creation of the Av map follows the procedure outlined in e.g. Prieto et al. (2014). In brief,  the map  is produced by taking the ratio of two images separated in wavelength. For NGC 1097, we take the ratio of  the HST images I-band and B-band. In such a ratio, regions depressed  by dust in the B-band are expected to get enhanced  at longer wavelengths, particularly in the I-band. The conversion of this image ratio to an  extinction map is done by   comparing I-band / B-band values   at locations in the   filaments and lanes, with  those  at dust-free locations  selected visually,   outside the ring.   Typically,   dust-free locations are selected at  distances no larger than $\sim $ 1.5 kpc radius  to minimise stellar  colour gradients. The Witt et al. (1992) extinction curve is used.  Fig. 1e shows the Av map for the nuclear region and the ring, Fig \ref{large}b shows the  large field-of-view   I-band / B-band image ratio  from which the Av map is constructed. 

The most conspicuous feature in the Av  map  - Fig. 1e - is the bright, tongue-like structure   that runs from the  Western side of the ring  to the  South of  it. It traces   
 one of the two major dust lanes of the galaxy, namely that  coming from North of  the galaxy (Fig \ref{large}),   at its entrance into the  ring.  The second  major dust lane   arises  South of the galaxy  and  as the Northern one, it  equally  extends from the edge of the galactic disc  straight to  the  ring. Yet, this South lane    is  not  showing up as prominently in the extinction map as  it   presumably enters the ring from the back, and the extinction map  traces dust  against a background emitting  source only.

The extinction  map further  shows  that most clusters   lie in  low dust regions, with  $A_V < \sim 1.5\, \rm{mag}$, in line with values derived independently from the stellar population analysis of the clusters  (sect. 4). The largest extinction, $A_V \sim 2 - 3$ mag, are found at the thickest filaments  and   at the prominent lane in the map. 
Extinction maps based on  other continuum image ratios,  namely  VLT/NACO Ks-band- and HST/WFC3 F814W-, F438W-, and F336W- images yielded consistent results, namely   $1.5 \lesssim A_V \lesssim 3 $ are found across the ring. This consistency is reassuring, it further     suggests the applicability   of the extinction  curve  used to     the central kpc  of  NGC 1097.

 A closer look at the conspicuous dust lane in Fig. 1e, or  at the larger scale view in Fig 3b,  shows that  it resolves into narrow, relatively long  filaments running parallel to each other  along the lane flow. The filaments can be  followed, continuously, from the lane through  the ring, and further in to the nuclear central parsecs. It can been seen that some of the filaments  as they enter into the ring at the South-East region of the ring they  follow the ring by circularising along the Southern region of it, others    curve immediately towards the  nucleus to produce the known nuclear spiral of filaments which can be  traced up to a few pc from the BH (Prieto et al. 2005).

To compare the extinction map with the molecular gas in the region, Fig. 1e shows the ALMA - HCN gas  at the same scale  on top of the Av map.  Albeit its lower angular resolution,  the HCN gas follows closely the location of the  dust lane inside the ring, where it circularises  at the Southern region, and  even outlines the   filaments at some locations. For example,     the HCN tongue-like feature  entering  the ring at the South-West region follows the  dust lane at that location, an equivalent HCN feature at the North-East  side of the ring  coincides with the entrance of the opposite  dust lane  into the ring.   North of the nuclear spiral, a correspondence  between an HCN feature bridging the ring and the nuclear  molecular gas is seen overlapping   with a  dust filament  that also crosses the ring and  curves into the nuclear spiral. These correspondences are equally seen in CO gas after comparing with e.g.  Izumi's et al. (2013) map.

\section{Discussion}
\subsection{Random age distribution in  the ring}

Clusters at all ages are found anywhere in the ring regardless of the evolutionary phase  i.e.  whether they are in the proto-cluster phase or already  formed. Moreover, clusters  do not group by   age or a special  location in the ring. Special locations  are  e.g., at the entrance of the dust lanes into the ring,  or  at the molecular-clouds location, being those places  where the younger clusters are expected to be.  

The  proto-clusters are the  only sources  that can certainly be associated with    the   molecular clouds as they  sit at the cloud peak emission  or close to it (Fig. 1). Because of this,  and  their moderate  temperature (sect. 4.2), they  may be very young,   perhaps  less than 0.5 Myr if put  in the context  of  the so-called
class 0 phase of  proto-stellar clouds,   these being known  as strong blackbody emitters in the far-IR (Evans et al.   2009; Seale et al. 2012).

There is marginal indication for the  youngest $\sim 4 $ Myr old population  to group slightly, in the sense that   there is a larger probability of finding a 4 Myr cluster next to a similar one,  but  otherwise  they  are largely mixed  in the ring, an indication  that  the clusters are moving away from their natal cloud  soon after their birth, in less than $\sim 3 - 4$ Myr  or, that   the natal clouds get destroyed by some kind of stellar feedback on short time scale. 
 
 The mixing of ages also   implies that clusters seen in projection on a given molecular cloud may not necessarily have arisen from that cloud, nor be representative of the star formation efficiency of that cloud, and therefore estimate of specific star formation per either cloud area or  a given section of the ring could be  misleading.

\subsection{Build up of the ring stellar mass }

The  amount of gas stored in the ring  is found  $\rm M_{\rm H_{2}} \sim 10^9\, \rm{M_\odot}$ (as derived   from the  HCN  gas, Hsieh et al. 2011).
Following on  the co-spatiallity between the  molecular-gas  and  the dust- lane and filaments at the entrance to the ring (sect. 5),  it is conceived that the  gas in the ring   proceeds and is replenished by  material flowing along  the two major  dust lanes   of the galaxy via  the dust filaments or streamers. We note that whereas some of the  dust filaments get through  the ring and inward,  others 
kind of     overshoot it, meaning that  they carry  a range of different specific angular momenta. Because of the large concentration of gas in the ring,  there should have  been over the past one dominant   component  of the momentum responsible for  the ring configuration     at the specific  ring radius that is observed.

Evidence for a net gas inflow along the large-scale dust lanes  of the galaxy  relies on  kinematic modelling of the  HI gas along   NGC 1097's bar by Ondrechen et al. (1989) who finds  a net inflow within the central 1.5 kpc radius.   Further in at the galactic centre,  a net  inflow  via the nuclear spiral of dust filaments  is supported by hydrodynamic and kinematic models (Prieto et al. 2005;  Davies et al. 2009; Fathi et al.  2013), with net    
inflow rates of  gas of $ \lesssim  0.6\, \rm{M_\odot \, yr^{-1}}$ from  distances of  100 pc and decreasing inward.

A first order  estimate of  the  filling factor  and  volume density of the streamers in the dust lane could be  derived  from the extinction map. A transversal cut at the North lane at  the position  marked in Fig \ref{large}b reveals its inner structure  in detail: the cut distinguishes up to three major bumps   - the  filaments - over a lane width of $\sim$ 280 pc, each  bump having  a width, FWHM, of $\sim 25$ - $30\, \rm{pc}$ (Fig \ref{large}c).  Presumably, these widths  are  upper limits set by the signal to noise in the lanes and our ability to spatially resolve the filaments in the lane. The filaments filling factor in the lane can be derived as the  ratio of the filament- to lane-section- areas, times the number of filaments in the lane section. Taking the lowest filament width,  $ff_{\rm lane} \sim 3~\rm{filaments} \times (25 / 280)^2 \sim 0.02$.  Note that this  is not a  filling factor per volume but per area as the lane is considered as a tube filled with filaments all running    parallel to each other  along the flow direction.

The filaments density  could be inferred from the extinction values. The highest extinction in the Av map is measured along the  dust lane  at its entrance into the ring,  $A_V~ \sim 2 - 3$ mag.  $A_V$ is converted to total gas  column density, $N_{\rm H}$, using the standard conversion $N_{\rm H} / A_V = 2 \times 10^{21}\, \rm{cm^{-2}\,mag^{-1}}$ (Savage \& Mathis 1979).  note that $N_{\rm H} $ traces  mostly molecular gas,  as HI is barely detected in the lanes (Ondrechen et al. 1989).
The volume density, $n_{\rm e}$,  is estimated from $N_{\rm H}$ assuming the filament depth equal to its width. Taken a conservative  average  width of 28 pc (see above), densities  in the range   $  n_{\rm e} =  50\, \rm{cm^{-3}} - 70\, \rm{cm^{-3}}$  are inferred.


Presumably, the source for this material running along the streamers is the  HI gas reservoir  of  $5\times 10^9 ~\rm M_\odot \,$, most of it   placed 
at the galaxy outskirts, as suggested by  Ondrechen et al. (1989). The  morphology of the HI gas embracing the NGC 1097 and its companion  NGC 1097 A further suggests the ultimate origin of this  gas  to be related to  the past interaction of NGC 1097 with its companion.

\subsubsection{Arrival mass through the streamers: feeding the ring and the BH}

A crude estimate of the arrival mass rate  through the streamers at the entrance to the ring  can be derived  from the   filament's  density and filling factor above estimated,   and assuming  a  geometry for the dust lane - we take it as cylinder with diameter, D $\sim 600\, \rm{pc}$, this is an  average lane FWHM  measured  in Fig \ref{large}b,  and a velocity for the gas flow, $v_{flow}$, at the entrance to the ring. For the latter, we use the fact  that    the filaments as they approach the ring they  circularise along it,  thus they are expected to  get coplanar with the ring  and share   the same specific angular momentum as that of the  gas in ring. The ring gas velocity is known   from the HCN kinematic to  follow  a  rotation curve that gets close to flat at  $v_{ring} \sim 350~ km~ s^{-1}$ (Hsieh et al 2011). We adopt this velocity   as that of  the flow at the entrance into the ring.  Using the expression:
 $  \dot{M}_{\rm entrance-ring} \sim D^{2} \times n_{\rm e} 
  \times ff_{\rm lane} \times v_{\rm flow}  \times 2 ~ _{dust-lanes} $, 
where the  factor 2 accounts for two dust lanes, and the  fiducial values above, a mass arrival rate $  \dot{M}_{\rm entrance-ring} $ $ \sim > 3\, \rm M_\odot \, yr^{-1}$ is found.

This is to be compared with the gas consumption in the ring and inward. The global current star formation in the ring estimated from the Pa$\alpha$ integration in the ring (see sect. 6.3) is $SFR = 1.8\, \rm M_\odot \, yr^{-1}$. In addition to this,     there is the material that following the   nuclear spiral  falls towards the centre, the estimated infall rate is    $\sim 0.6 \, \rm\
M_\odot \, yr^{-1}$ at 100 pc from the BH  (Fathi et al. 2013). An upper limit to the BH accretion rate  as derived from the central 10 pc SED (sect. 1) is orders of magnitude less, $ 10^{-5}  \, \rm{M_\odot \, yr^{-1}}$  using a 10\% conversion efficiency. Summing up the above numbers, the estimated gas consumption  in the central kpc is   $ \lesssim 2.4\, \rm{M_\odot \, yr^{-1}}$. Subject to the approximations made,  this consumption may comfortably be accounted for by the streamers.

%
\subsection{Star formation regulation in the ring}

The  star formation rate  (SFR) in the ring  as derived from the integration of the Pa$\alpha$ emission,   after extinction-correction using  the H$\alpha$/Pa$\alpha$ extinction map (Fig. 1), and converted to a star formation rate using  Kennicutt \& Evans'  $H\alpha$ calibration (2012) yields  
$SFR = 1.8\, \rm{M_\odot \, yr^{-1}} \times ~F({\rm Pa\alpha})_{\rm corr} / 1.6 \times 10^{-12}\, \rm{erg\,cm^{-2}\,s^{-1}}$, where  $F({\rm Pa\alpha}_{\rm corr})$ is the $F(\rm Pa\alpha)$ flux  corrected for extinction.
This   SFR   is a factor 2.5 lower than previous estimate by Hummel et al. (1987), the difference being   due to the higher, by same factor $\sim$ 2.5,  SFR calibration factor applied by those authors. 

The  gas depletion timescale,  i.e., the ratio of molecular gas mass to SFR,   is  $M(H_2) / SFR $ =  $10^9\, \rm{M_\odot} /  1.8\, \rm{M_\odot \, yr^{-1}} \sim 5\times 10^8 yr $,  in line  with  the universal depletion timescale of molecular gas in star forming galaxies, in the Gyr range (Tacconi et al. 2018 and references therein).
Yet, the  total    mass  in   clusters in the  ring  older than say,  20 Myr  and up to 100 Myr, the age of the older clusters dated in the ring, is found to be   
$  \lesssim 8 \times 10^6\, \rm{M_\odot}$  (cluster's mass is derived from the SSP analysis, sect. 4). At  the current SFR   of $ 1.8 ~{\rm M_\odot \, yr^{-1}} $,  the ring should have produced   a total  of $\sim 2 \times  10^8\, \rm{M_\odot}$ in clusters mass instead, i.e   an order of magnitude  more in stellar mass.  It thus  appears that factors other than the  availability of  gas  are more determinant in  the regulation of star formation,    perhaps    the  rate at which   dense gas accumulates in the ring to become  self gravitating for  star formation to begin     is more crucial (Gao \& Solomon 2004; Burkert \& Hartmann 2013; Burkert 2017; Evans 2017).

The ring is a place of bursty, recurrent star formation.
The intermittent star formation may be due to an
interrupted feeding or perhaps to   conditions  in the ring unsuitable  for star formation. If the ring is fed by the streamers, the  collimated morphology of the lanes
over several kpc distance across the galaxy  and the large  concentration
of gas in the ring suggest  a rather continuous
flow of matter through the lanes (still, the kinematics of HI along the bar indicates a
more complex dynamic, Ondrechen et al. 1989 ).

Here we focus on  a more local process, namely the   suitability of   the gas in the ring  to form stars and examine  the  Toomre parameter, Q, which is a measurement   of the stability of a disc of gas  against collapse.  For an infinitesimal thin disc, Q =1 indicates a stable disc,  but for thick discs, stability still remains  for Q $\sim>$ 0.7 (e.g. Behrendt et al. 2015).  Following Toomre (1964):

$Q \sim c_s ~ \kappa / \pi ~G ~ \sum_{\rm gas-density}$
\\
where $c_s$ is the sound speed, $\kappa$, the epicyclic frequency,  is approximated as  $ \rm v_{\rm ring} / \rm radius_{\rm ring} $, G is the gravitational constant,  $\sum_{\rm gas-density}$ is the ring surface gas density.

 Typical temperatures of the diffuse gas in the  Milky-Way are  $T_{gas} \sim 3000 - 10,000K$ (e.g.  Kulkarni \& Heiles, 1998) yields  $c_s = 6 -10~ \rm km ~\rm s^{-1}$;  the gas in the ring is rotating with $v_{\rm ring} \sim 350~\rm km~ \rm s^{-1}$ (Hsieh et al. 2011),  the total gas mass is  $\sim ~10^9 \rm{M_\odot}$ and the  internal  and external ring-radii are  taken 600 pc and 1 kpc respectively. Substituting values results in   Q  $\sim$  0. 9 - 0.6,    pressure terms as  turbulence  or/and magnetic field ( e.g. Tabatabaei et al. 2017) are not even  considered, meaning   that  the ring is presumably     in  a marginally stable regime.  It is thus possible that in the past,   may be roughly the last  few 100 Myr, the streamers have fed the ring without forming (a significant amount of) stars because at that time  the disc  was stable against collapse. Yet, at some point, enough mass had accumulated such that the increasing surface density pushes Q below one and stars  start to form. It is thought to be  some kind of self-regulation so that  Q  remains close to one (Dekel et al. 2009; Burkert et al. 2010): once enough gas is accumulated,  star formation  starts, gas is consumed and Q is pushed above one again.   Such scenario   could  explain the small mass in stars and  intermittent star formation in the ring.




\section{Summary: the overall view}

This work presents a  multiwavelength, parsecs-scale study  of the young cluster population and  the interstellar medium enshrouding it, dust, ionised and molecular gas, in the  kpc-radius circumnuclear ring of the galaxy NGC 1097.
Circa 250 young stellar clusters are individually resolved in the ring. About $ 70\%$  show emission up to the UV, thus avoiding   the numerous dust filaments that cross the ring.  The remaining group, hidden by dust, is recovered at near-IR wavelengths, chiefly in the K-band (sect. 3, Fig.1).  The clusters' properties fit entirely within those of  massive star clusters in the MW and nearby galaxies: masses in the $10^4 -10^5 \, \rm{M_\odot}$ range, ages in the Myr to a few 100 Myr range,  luminosities in the $10^2  - 10^3~ {L_\odot}$ range. Their sizes are however limited by the angular resolution of the data,  FWHM$< 8~ pc $, their  extinctions inferred from different methods  are moderate, $A_V < \sim 2\, \rm{mag}$ (sect. 4).
 
The ring is  a place of intermittent bursts of star formation spreading over the last few 100 Myr. Several major  burst episodes   are identified, these   cover  a proto-cluster phase,  a major one   at $4 \pm1$ Myr, and older ones       at  $\sim 30 \pm 10 \rm Myr$, and  possible one, perhaps two  episodes   between 50 Myr and 60 Myr and in between 80 Myr and 100 Myr. All together,  they may separate by an elapsed time of 20 - 30 Myr (sect. 4.1, Fig \ref{bursts}). 
  
  The proto-cluster  phase comprises  about eleven $11$--$18\, \rm{\micron}$ point-like sources,   sizes $FWHM < 60\, \rm{pc}$,  strategically placed at  the core   of the  HCN (J=1-0)  molecular clouds in the ring, which are  FWHM $\sim$ 100 pc in size (Fig. 1). Their emission is in all cases consistent with a $\lesssim 150\, \rm{K}$ grey black-body temperature, and they are shielded from the external radiation by thick envelopes  that represent extinctions Av in the 50 - 80 mag range (sect. 4.2). These sources   may presumably be cluster factories,  possibly still in the cooling phase.  Judging from the older bursts in the ring, taking as an  average rate   of about  20 to 30 clusters per burst  (Fig \ref{bursts}), each of these proto-clusters should  give birth to two to three clusters each.

 Cluster of any age  are found at all locations  in the ring, old and young  mix at all positions. The proto-cluster  population also  spreads all over the ring. The mixing may be caused by the incoming  of    material through the two major streamers into the ring, the critical time scale  for mixing    being  (sect. 6.3)
 $1/2 (2 \pi \times {\rm ring-radius}  ~/~ v_{\rm ring}) \sim  5$ Myr, the time needed for the gas to move from one streamer to the   opposite one in the ring.  This timing may conmesurates  with  the onset of the pre-cluster population seen  at all  positions in the ring. 


Judging by the  large proportion of clusters  detected up to the UV, their  clean environment  is presumably a consequence of the onset of winds by new born stars  which  sweep out the parental material from which they form.  The cleaning  should happen at the early phases considering that  the youngest 4 Myr population  is mostly dust-free (Fig. 1). 

The ring has a molecular mass of $M_{H_{2}} \sim 10^9\, \rm{M_\odot}$. It  produces stars at a current rate of SFR = $1.8\, \rm{M_\odot \,yr^{-1}}$ as inferred from the Pa$\alpha$ gas (sect. 6.3), and there is an  additional drain of material   to the galaxy centre of $0.6\, \rm{M_\odot \,yr^{-1}}$  through the nuclear spiral filaments. We believe  all this  material  
is being supplied by  the two major  dust-gas streamers that flow along NGC 1097 bar to the ring, the source of this material being as suggested by  Ondrechen et al. (1989), the envelope of $10^9 \, \rm{M_\odot}$ HI gas  at the galaxy  outskirts. The streamers    resolve in narrow,  FWHM $< 25~ \rm pc$, filaments that are followed from the dust lane into the ring with  undisrupted morphology. As they approach the ring, they start circularising along it,   some continue further in circularising long the ring, others curve inward to produce the  nuclear spiral (Fig \ref{large}b). The filaments  fill the lane, across the transversal direction with a filling factor of   about $ 2\%$.  A first-order estimate of the arrival mass rate through the streamers at the entrance to the ring is found  to   be largely sufficient   to account for the total  current global SFR in the ring and infall mass rate to the galactic centre (sect. 6.2.1).

  
 The  gas depletion timescale in the ring, $M(H_2) / SFR  \sim 5\times 10^8 \rm yr $, is within the range of that observed in general in starforming galaxies. Yet, the total  mass    in    clusters in the  ring  formed over the last  100 Myr  is $\lesssim8 \times 10^6 \, \rm{M_\odot} \ $ one order of magnitude at least, below of what is expected on the basis of the current   SFR.  It thus appears  that the  star formation in the ring is  independent of the  gas at disposal,    perhaps    the rate  at which  dense gas accumulates at the clouds is more critical.  The stability criterium of gas in the ring, following  Toomre parameter,   yields $Q >  0.6$ without considering any pressure term due to e.g. turbulence, which we interpret as the disc    being in a marginally stable regime. It is thus plausible that over the last   100 Myr, the streamers have fed the ring without forming (a significant) amount of stars because the  gas density may not have reached the threshold  for disc fragmentation, i.e. Toomre parameter was much larger than one.
Yet, at some point, enough mass had accumulated such that the increasing  surface density pushes Q below one and star formation begins. 
A first order estimate of individual SFR for   the older bursts in the ring points to a low efficiency  process. Assuming a burst duration of 2 Myr (judging from the age dispersion of the 4 Myr burst), an average mass per a cluster of $5\times 10^4 \, \rm{M_\odot}$ and an average     of  20 - 30 clusters  per a  burst yield  SFR between  0.7 to  0.5 $\, \rm{M_\odot} ~ \rm yr^{-1}$ per burst, i.e. more than a factor two below the current one.

\section{Acknowledgements}
AP thanks the CAST group of the Ludwig Maximilians University of M\"unchen (LMU) for the numerous comments and  their enthusiasm,  and the Max-Planck Institute f\"ur extraterrestrische Physik for the continuous hospitality.  Thanks to T. Naab, J. M. Vilchez, C. Mu\~noz-Tu\~non, G. Tenorio-Tagle, for discussions.


\begin{figure*} 
\includegraphics[]{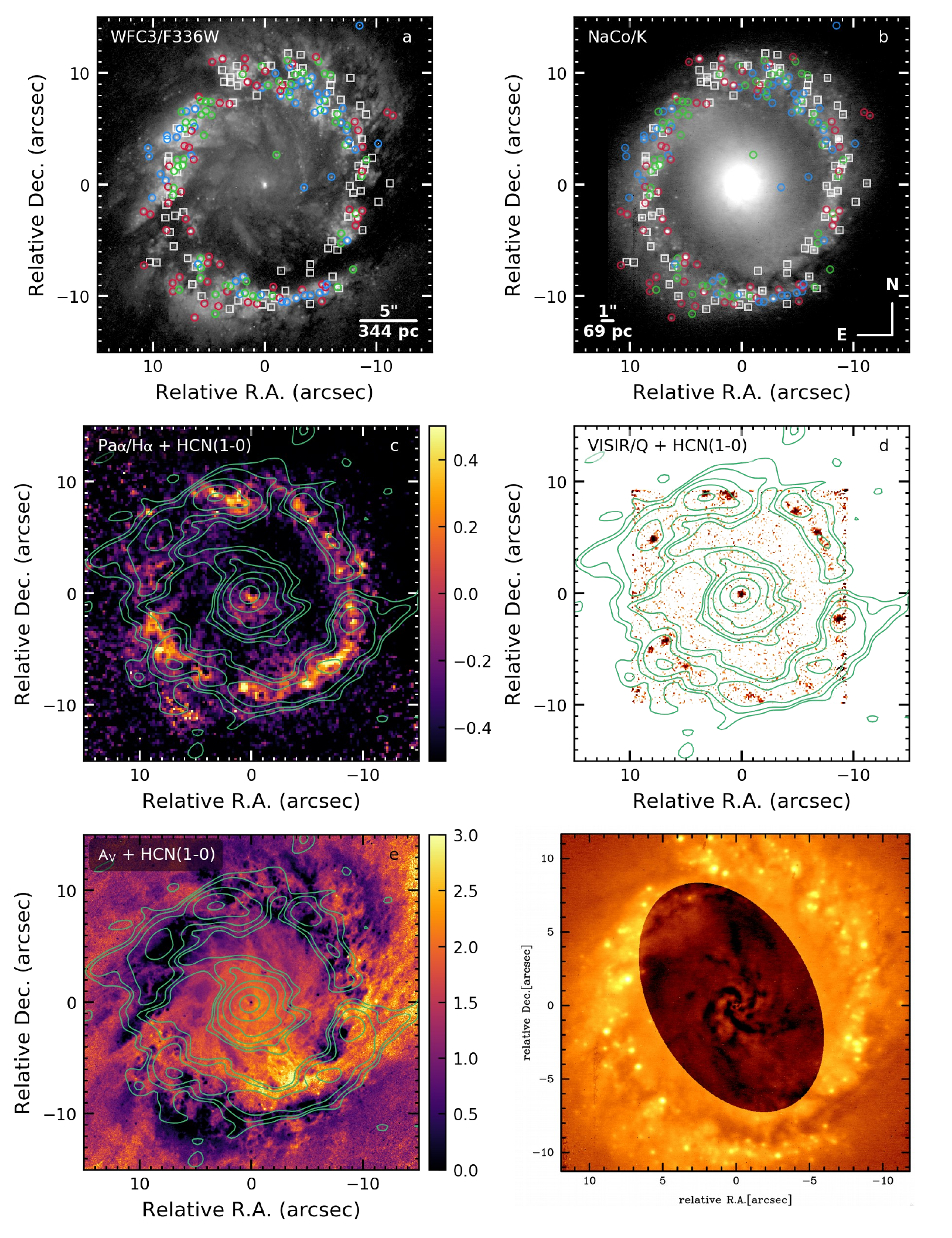}
\caption{NGC 1097 central $2\, \rm{kpc} \times 2\, \rm{kpc}$.  a: HST/ F330W image, cluster ages are colour coded: blue, $ < 10\, \rm{Myr}$, green, between  10 Myr to $40\, \rm{Myr}$; red, $> 40$ Myr to a few $100\, \rm{Myr}$, white squares are clusters detected longward $0.8\, \rm{\micron}$ only and not dated (see text); b:  VLT/NACO Ks-band with colour code as above; c: $\rm Pa\alpha/H\alpha$ map in log scale (log  $\rm P\alpha/H\alpha_{\rm case-B}$ =  - 0.94) with ALMA / HCN (1-0) in  contours;   d: VLT-VISIR $18\, \rm{\micron}$  image, the proto-clusters are the point-like reddish sources close to the peak emission of the HCN(1-0) gas in contours; e: Extinction map Av inferred from HST image ratio F814W / F438W with  HCN(1-0) in contours; f: VLT/NACO J-band image, the oval shape is a sharp, enhanced image of the nuclear spiral after removing  the galaxy light  (adapted from Prieto et al. 2005). g: X-ray Chandra image in contours on top of HST/ F330W image. In all panels, 0",0" is the reference location for the nucleus in NACO / K-band. Relative astrometry between all the images is better than 30 mas, angular resolution FWHM $\lesssim$ 0.15  arcsec  (sect. 2).}\label{clusters}


\end{figure*}
\begin{figure*}
\includegraphics[width=0.5\textwidth]{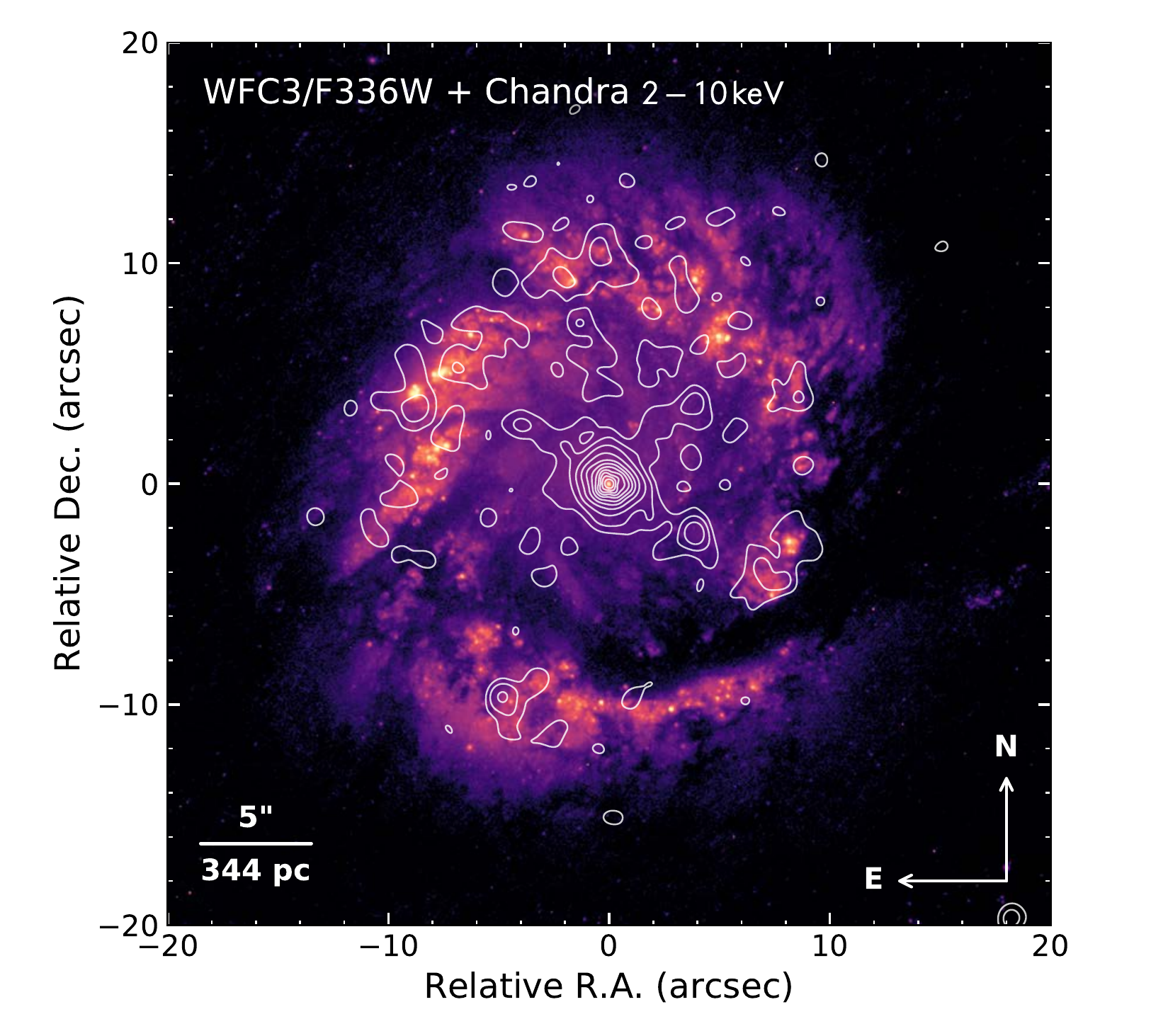} 
\contcaption{}
\end{figure*}

\begin{figure*}
\includegraphics[width=1\textwidth]{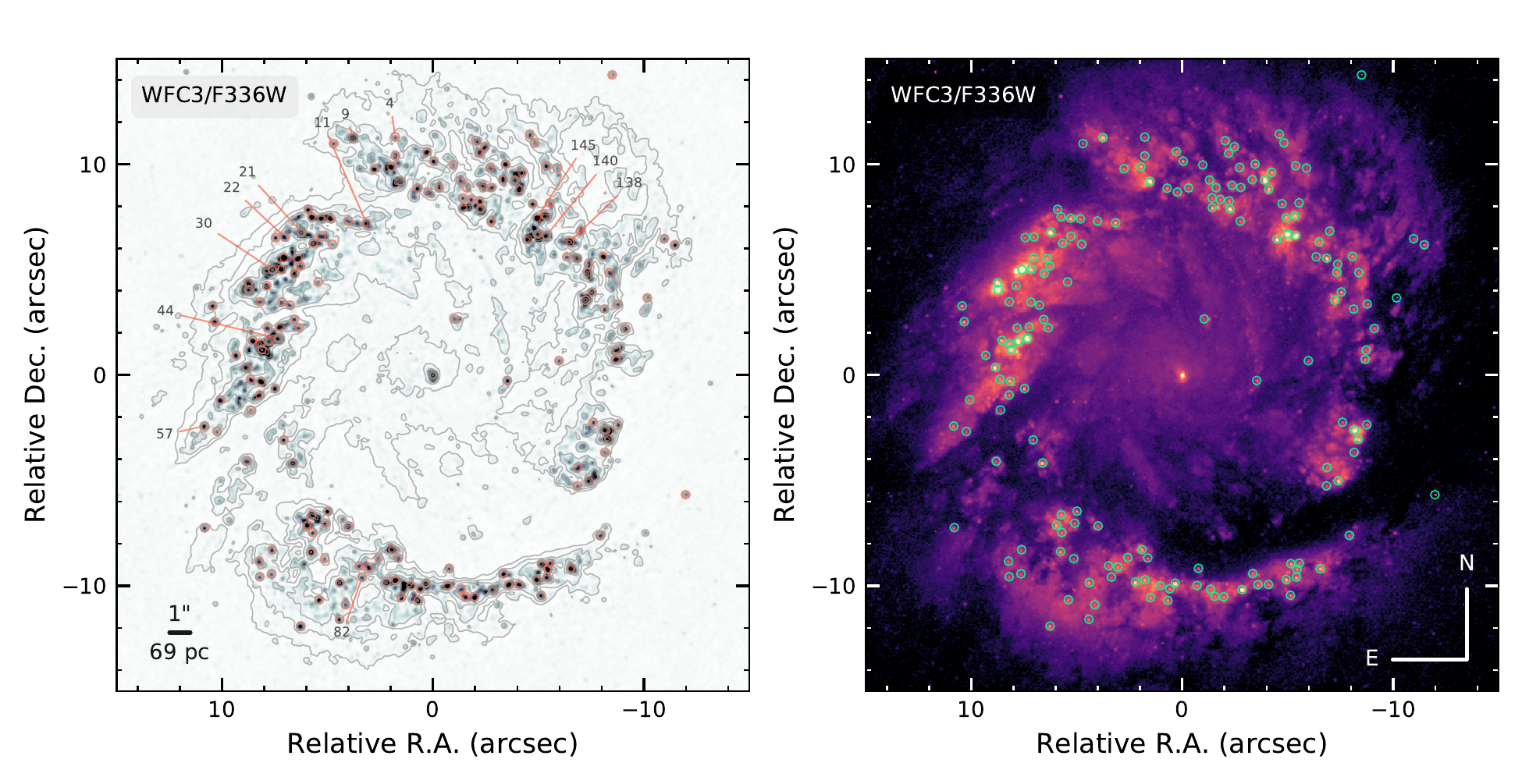}
\includegraphics[width=1\textwidth]{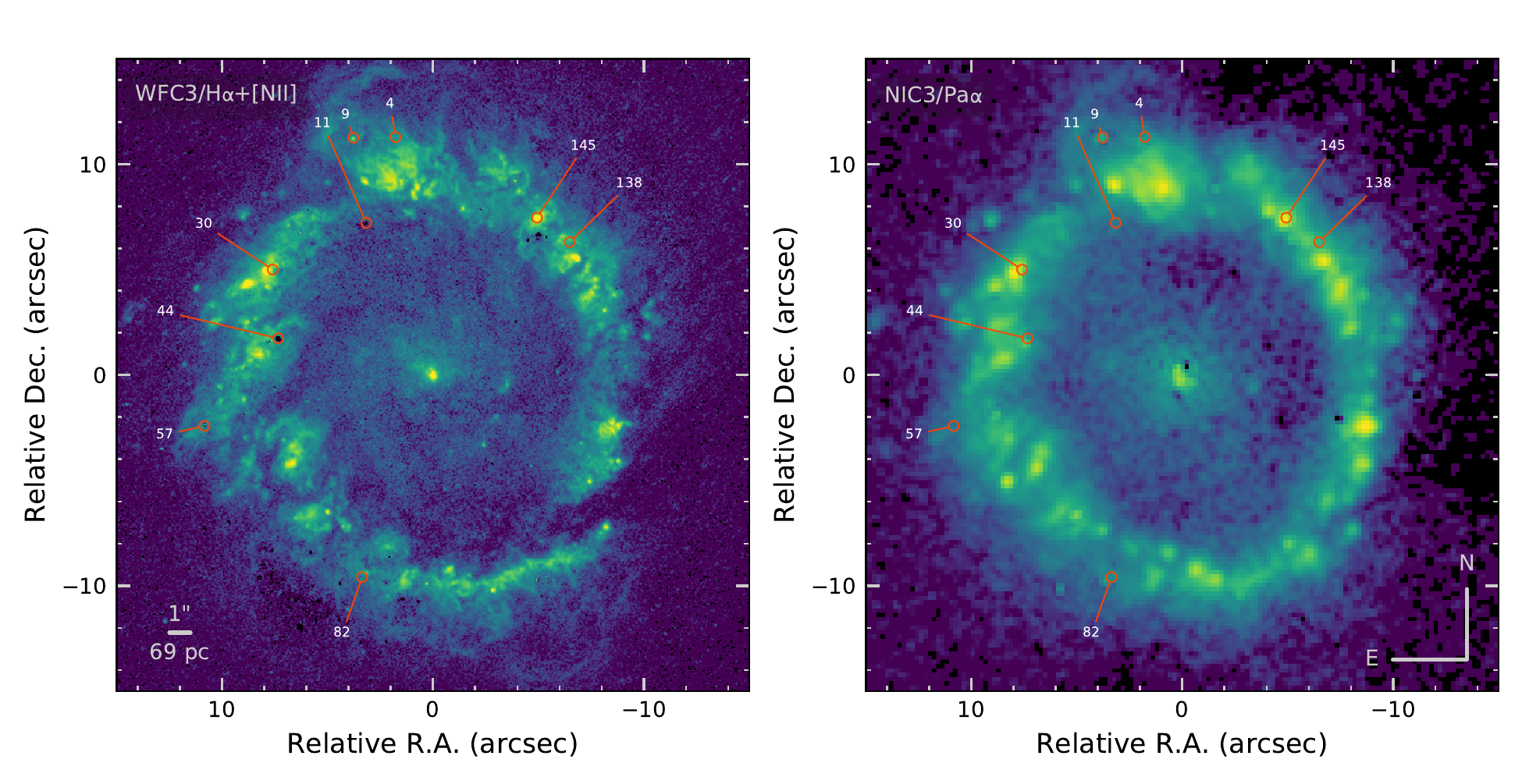}
\caption{To best illustrate the clusters both identification and  strength,  top-left panel is the HST-WFPC3 UV image in contours after the galaxy light being removed, top right panel shows the original image,  same as that in Fig.1. Circles in both panels enclose the clusters identified in the UV only  that passed a three sigma detection limit, 171.  SSP fits were applied to this set  only (sect. 3). A selection of  clusters from the main age periods and range of Av   identified in  the ring are marked in the top left panel. SSP fits for those    are shown in Fig. \ref{fit}. 
Lower panel: Left is the HST-WFPC3 H$\alpha + \rm  [NII]$ image; Right: HST-NICMOS Pa$\alpha$ image. Both have  the continuum subtracted (sects. 2 \& 3.2). Note that Pa$\alpha$ image is about a factor three lower in angular resolution than the H$\alpha + \rm  [NII]$ image. Identified clusters are the same as those in the top panel. }\label{ha}
\end{figure*}

\begin{figure*}
\includegraphics[width=0.5\textwidth]{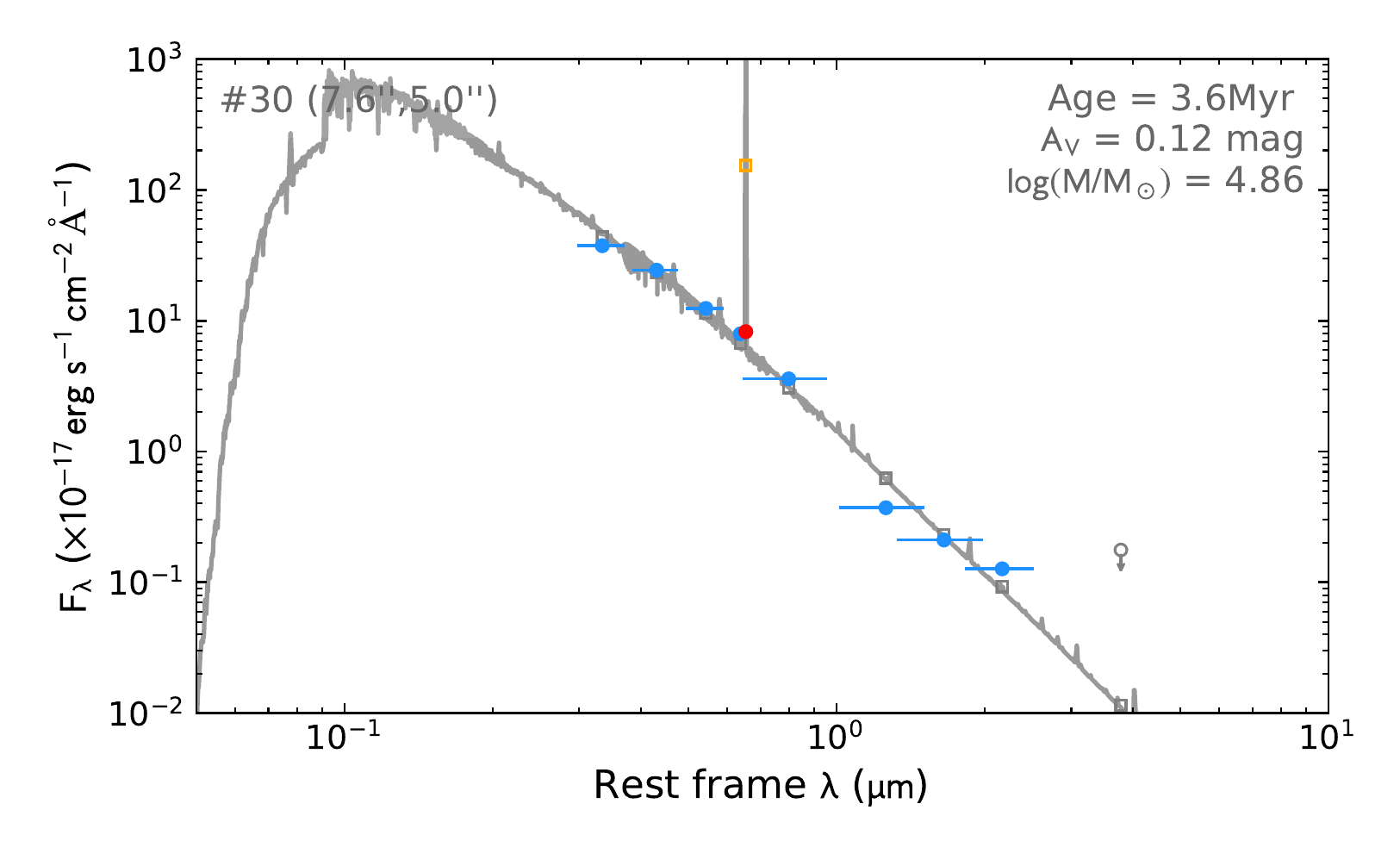}~
\includegraphics[width=0.5\textwidth]{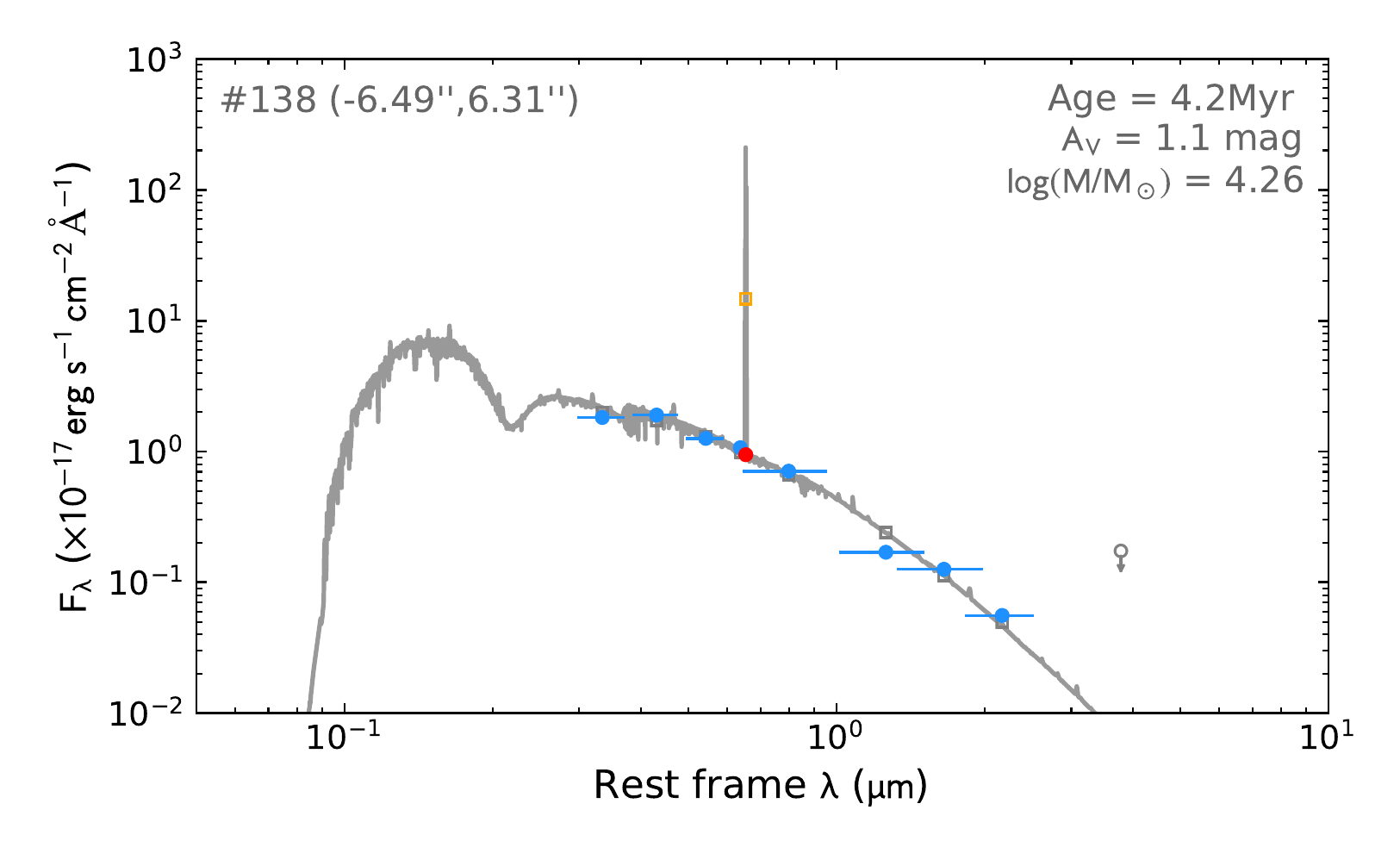}
\includegraphics[width=0.5\textwidth]{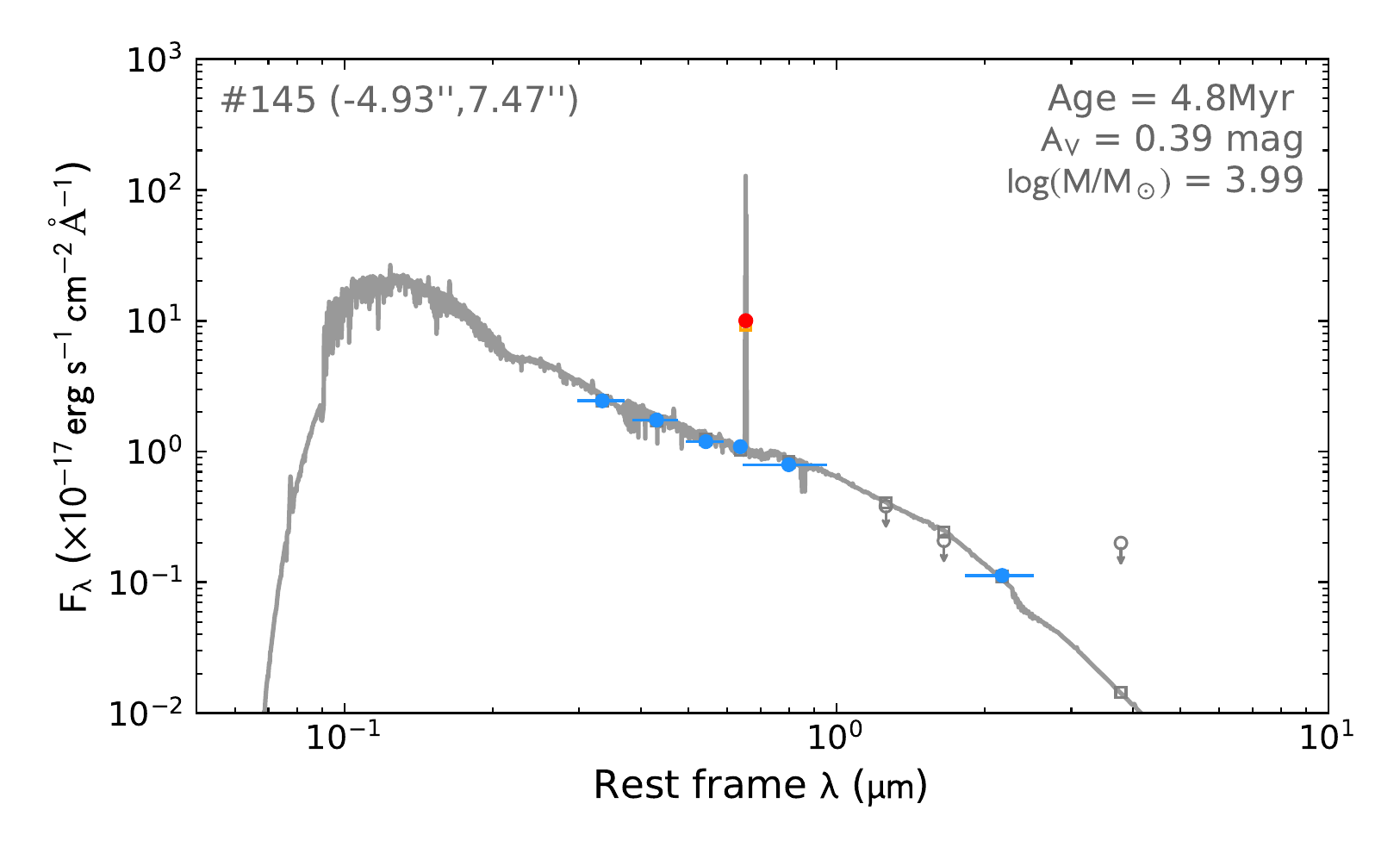}~
\includegraphics[width=0.5\textwidth]{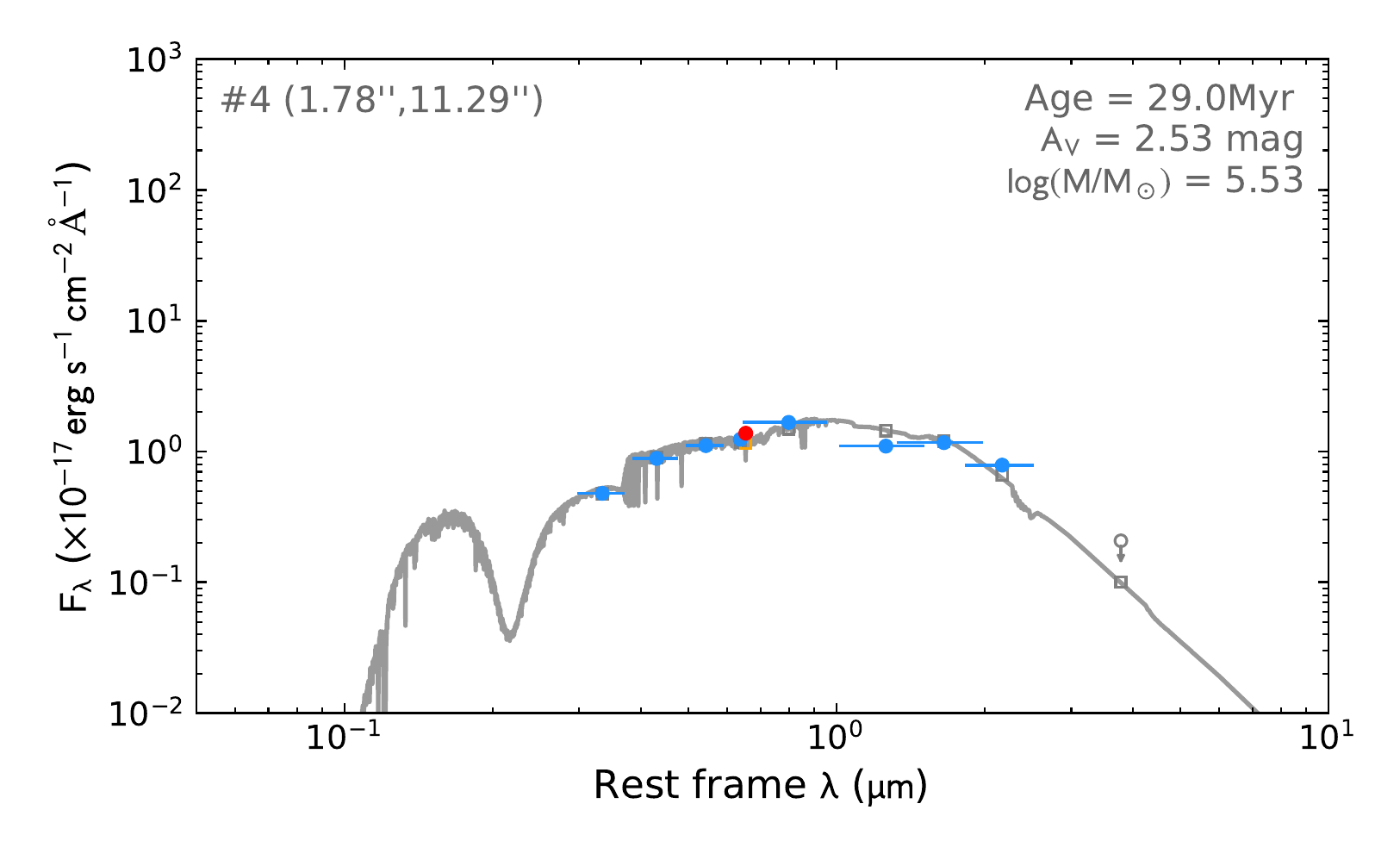}
\includegraphics[width=0.5\textwidth]{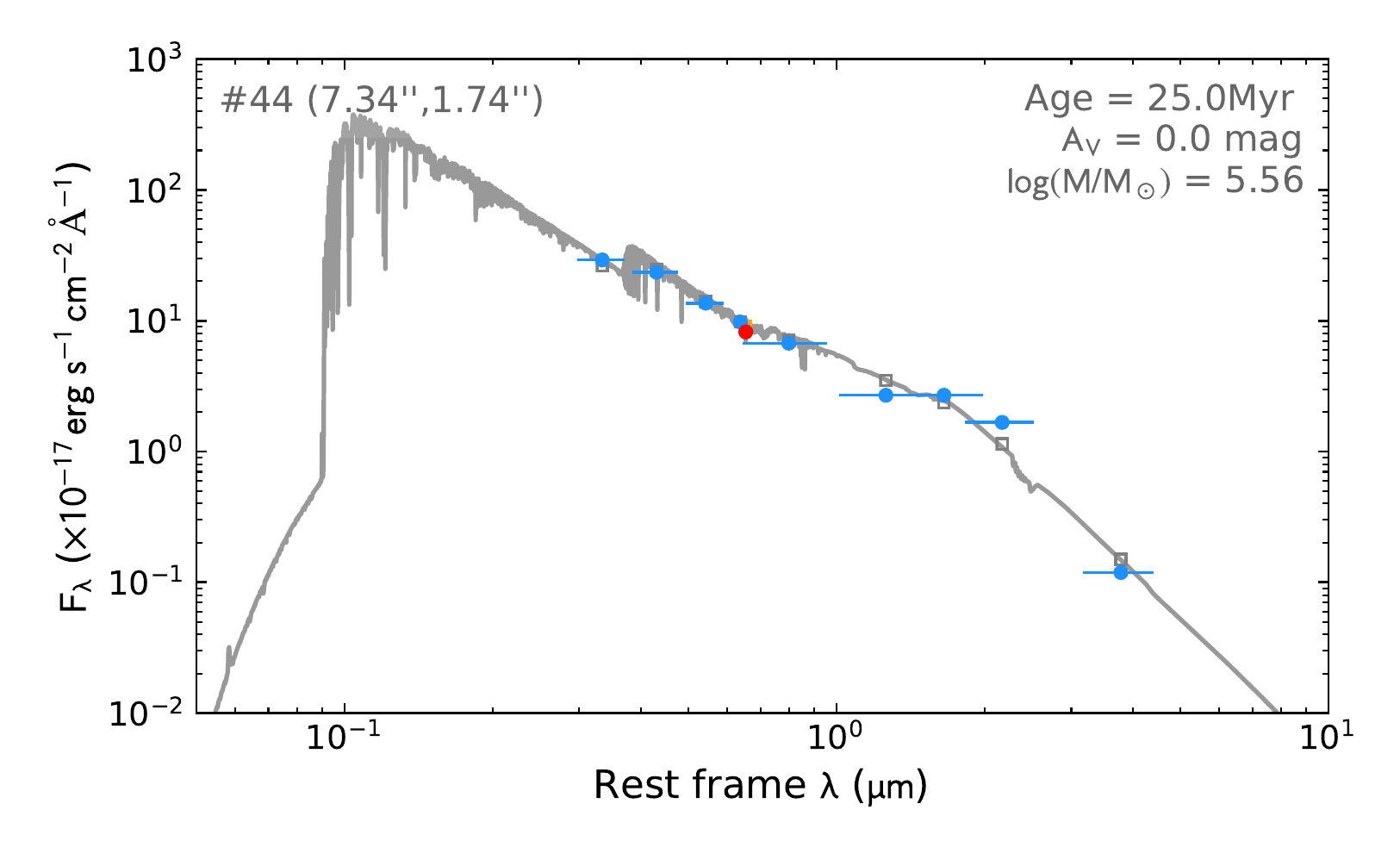}~
\includegraphics[width=0.5\textwidth]{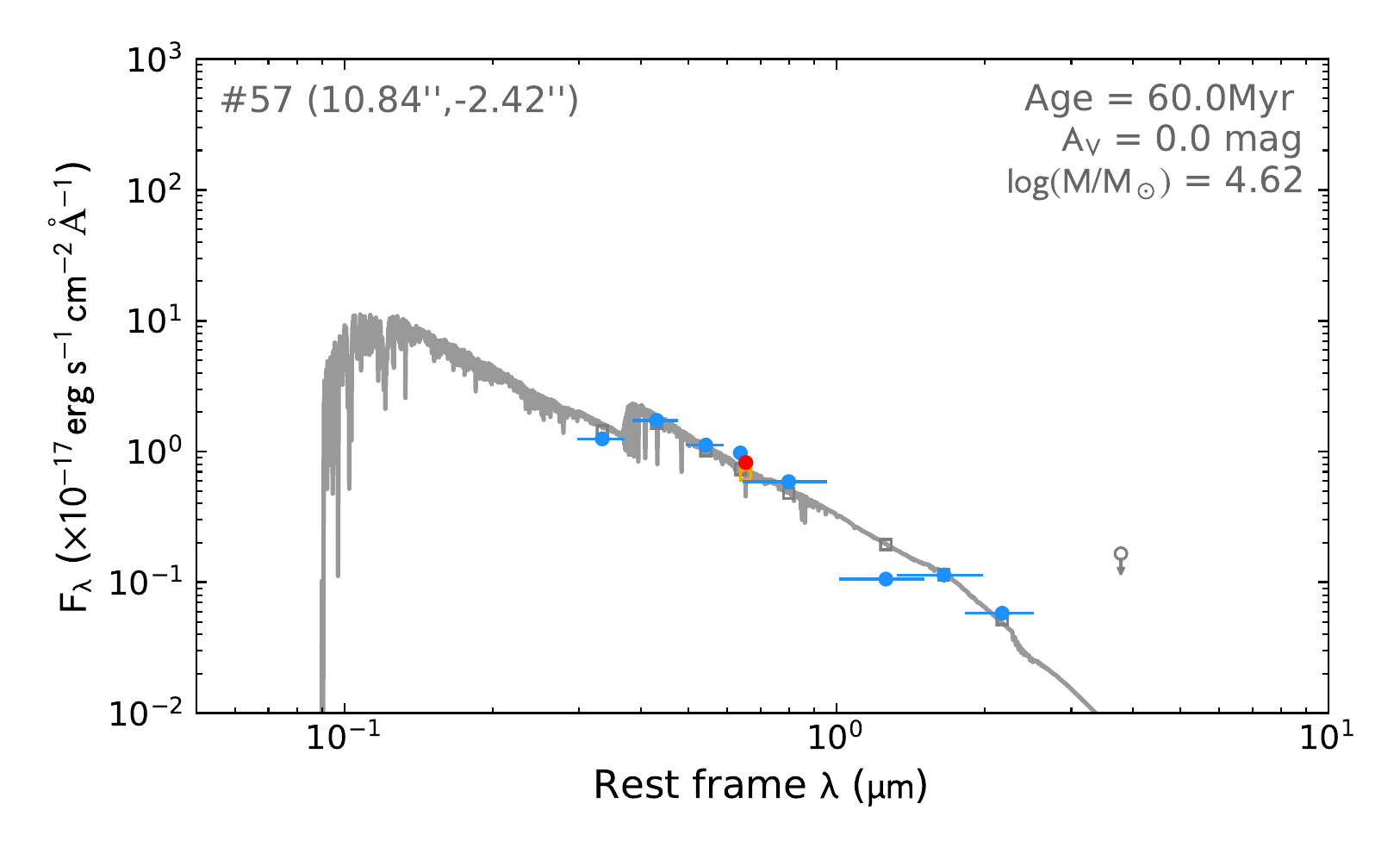}
\caption{Examples of SSP fits to a  representative set of clusters selected from the main age periods in the ring. Age, mass and Av inferred from the fit are indicated. Data are circles,  model is the line,   squares are the model after convolution with  the filter width used for the data, the horizontal bar indicates the width of the filter. The red point  is the photometric emission  in the HST / Ha+[NII] filter, which includes the line and continuum  emission; the yellow point on  the vertical grey line  is the predicted H$\alpha$ emission from the  ionising budget of the cluster and added up to the continuum prediction, which includes the photosphere contribution, all  convolved with the shape of the HST filter (hence the grey line in the younger clusters). 
The cluster number and its relative coordinates -  left corner of each plot - are to  be used to find its location in any of  the images in this work. To illustrate the  range of extinctions and its effect,   per each age period  two distinct  clusters, one requiring minimum Av, other requiring  larger Av,  are shown. 
It can be notice the discrepancy between the predicted $H\alpha$ emission and the data for the youngest 4 Myr burst (illustrated with the  clusters 30 y 138), which is due to  the diffuse nature of the nebular  gas in the ring (discussed in sect. 3).  The cluster no 145 is an example of the few in which the measured emission is in line with the prediction.}\label{fit}
\end{figure*}
\begin{figure*}
\includegraphics[width=0.5\textwidth]{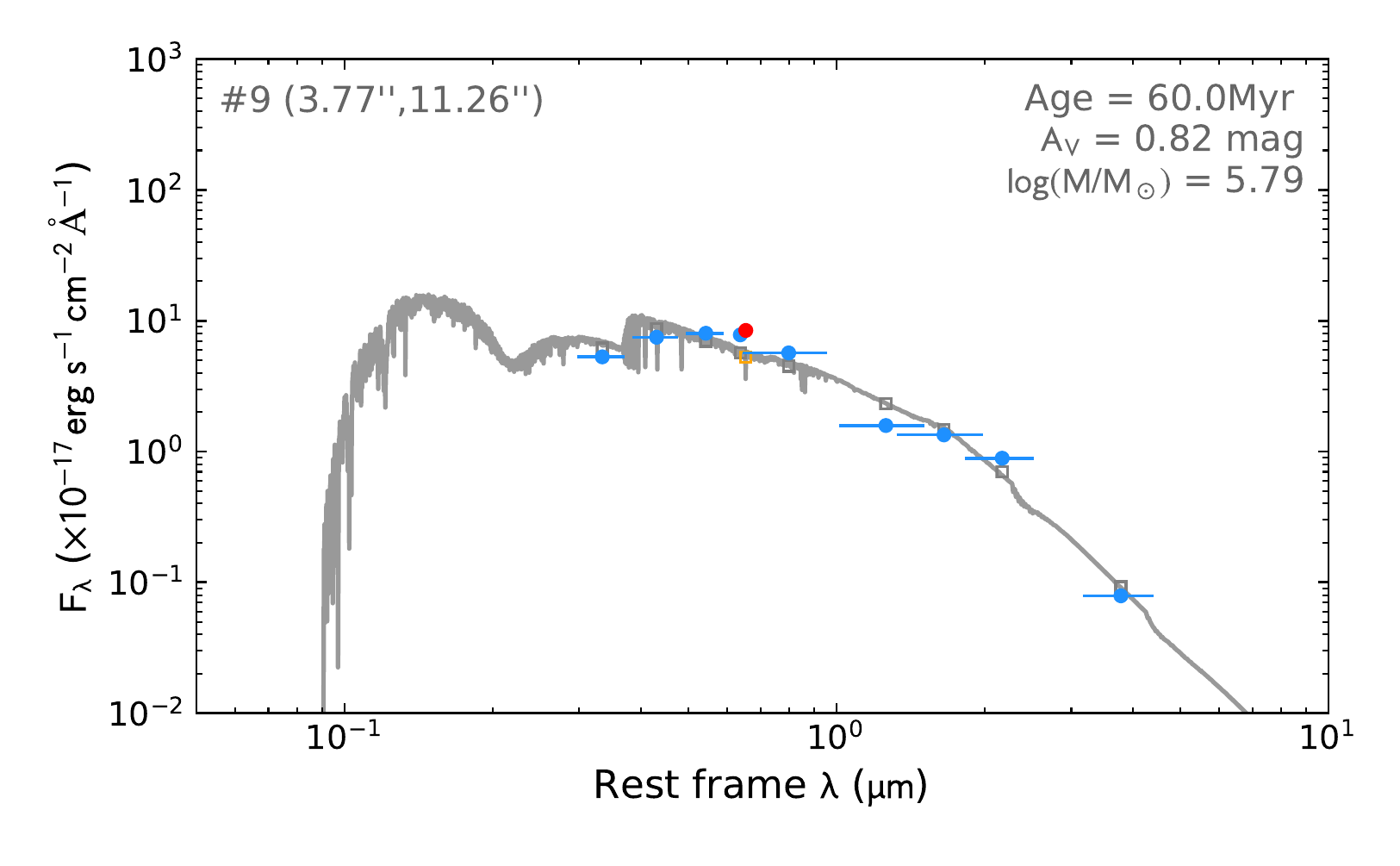}~
\includegraphics[width=0.5\textwidth]{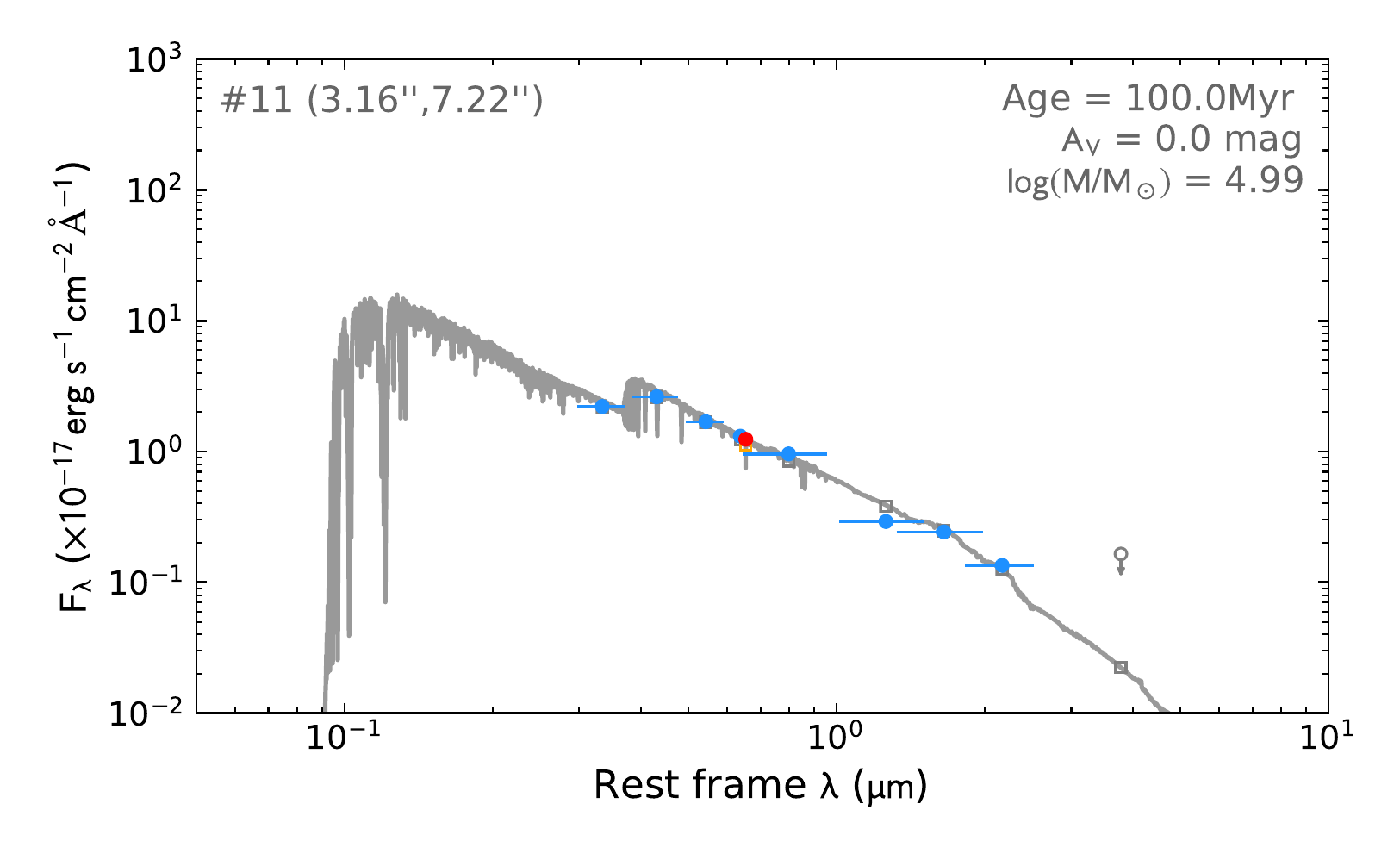}
\includegraphics[width=0.5\textwidth]{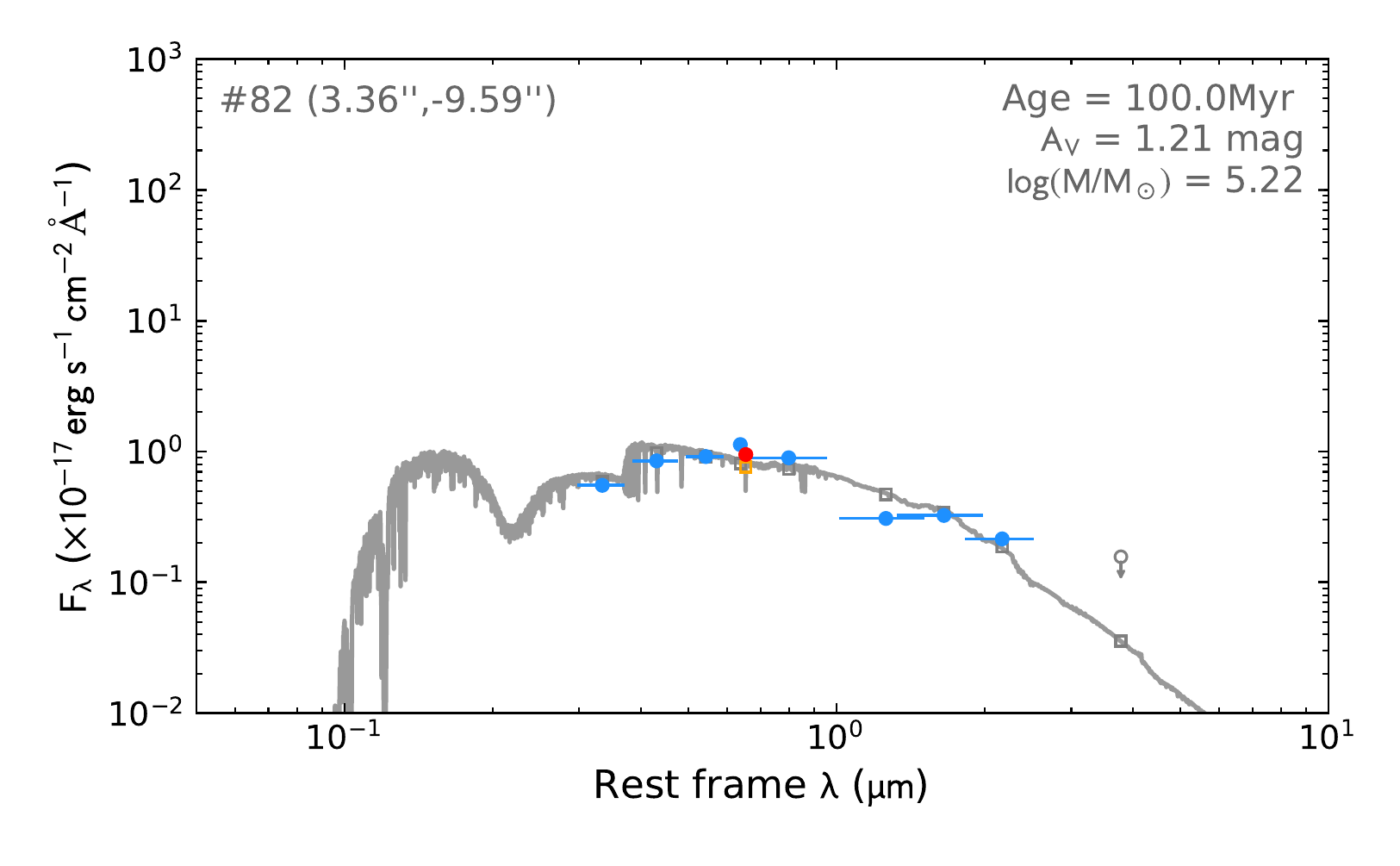}
\contcaption{}
\end{figure*}

\begin{figure*}
\centering
 \includegraphics[width=0.8\textwidth]{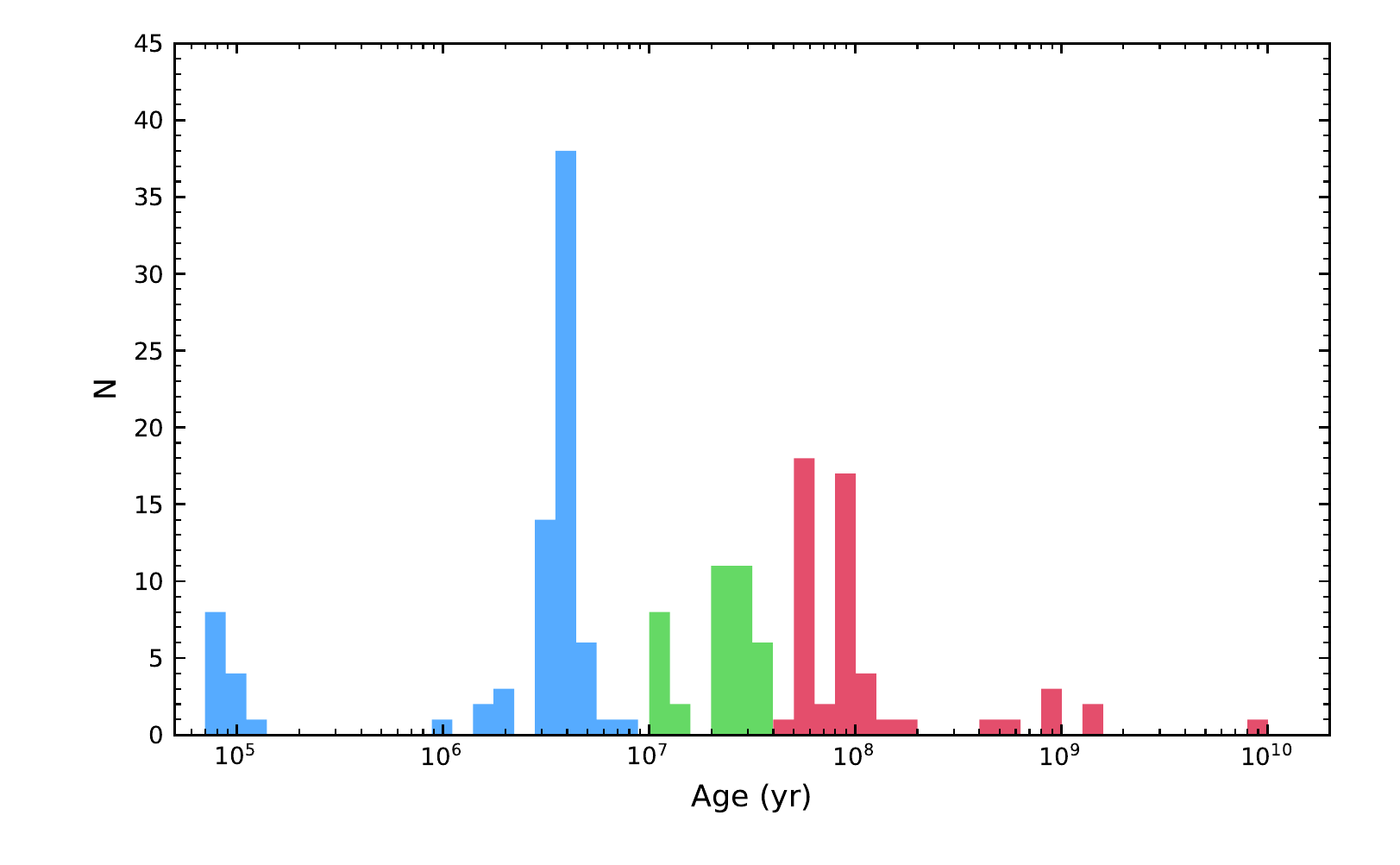}
\includegraphics[width=0.8\textwidth]{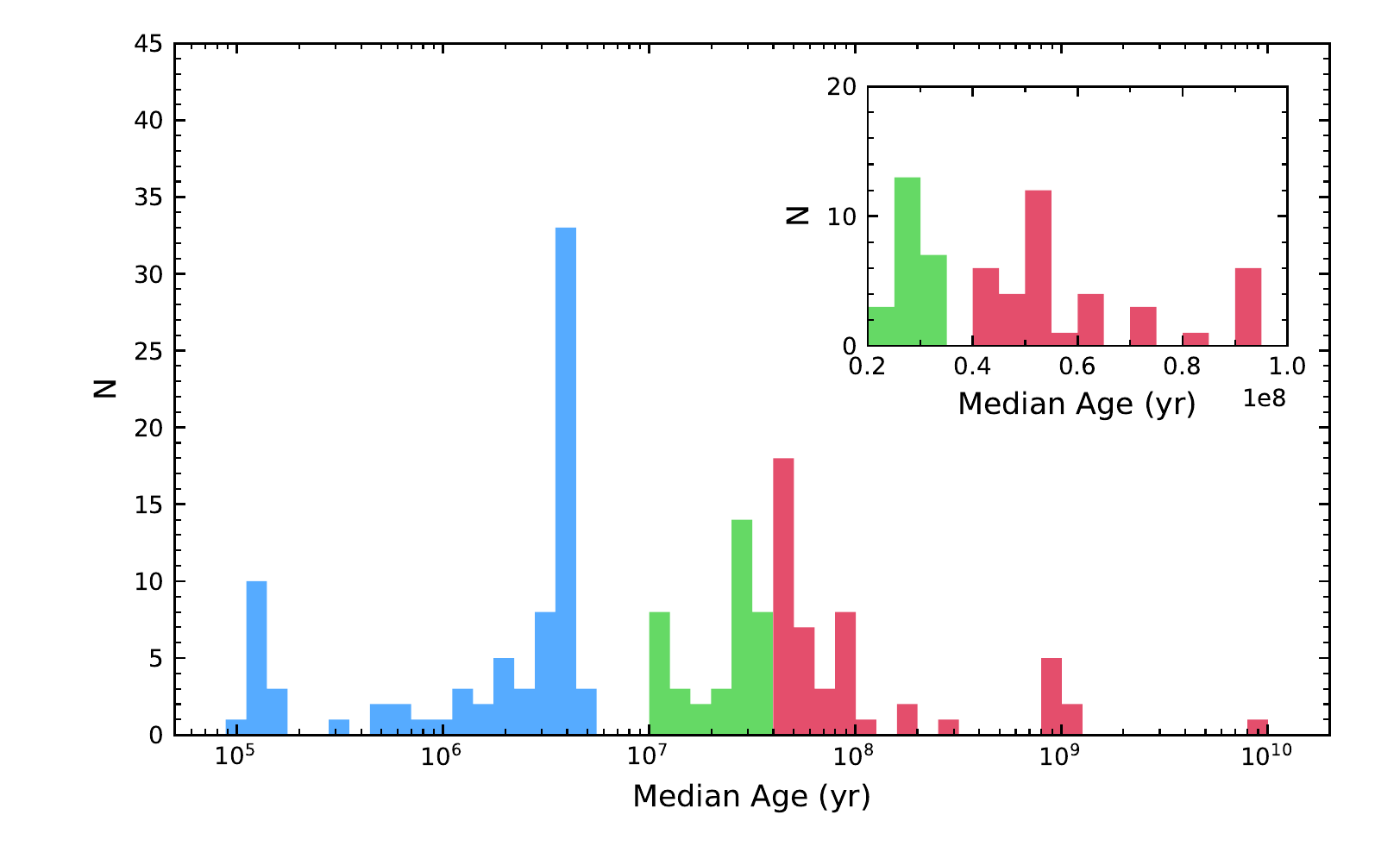}
\caption{ Histogram of ages  in the ring for the 171 clusters with secured  UV detection. 2a: Ages result from the best SSP fit to the SED of the cluster following a $\chi^2$ minimisation;  2b:  the age assigned to each cluster is the median of the 10  best $\chi^2$ fits to its SED; the purpose of this histogram  is to asses the age uncertainty of  three - four major  star burst periods identified in the ring.}\label{bursts}
\end{figure*}

\begin{figure*}
\centering
\includegraphics[width=0.4\textwidth]{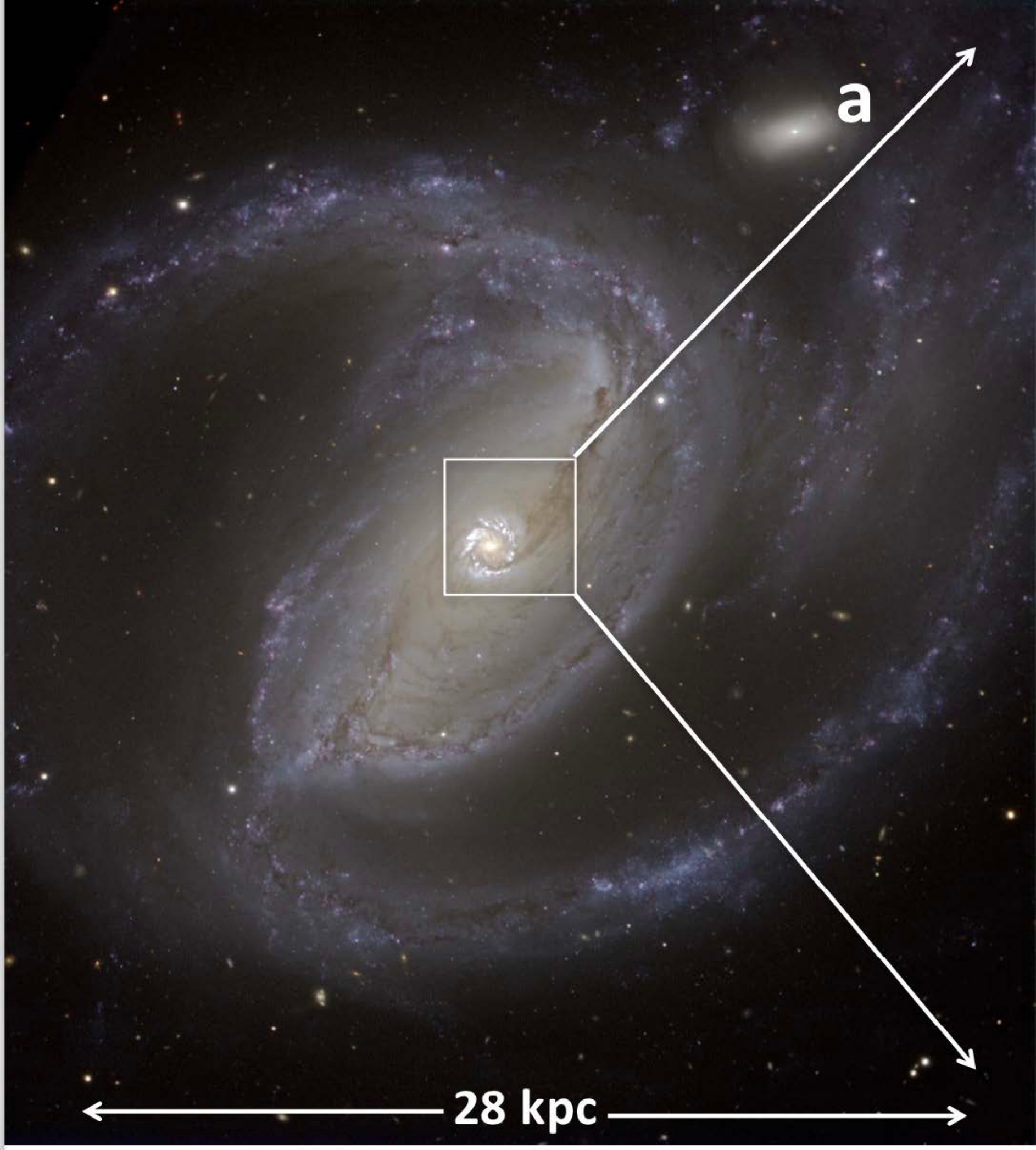}
\includegraphics[width=0.48\textwidth]{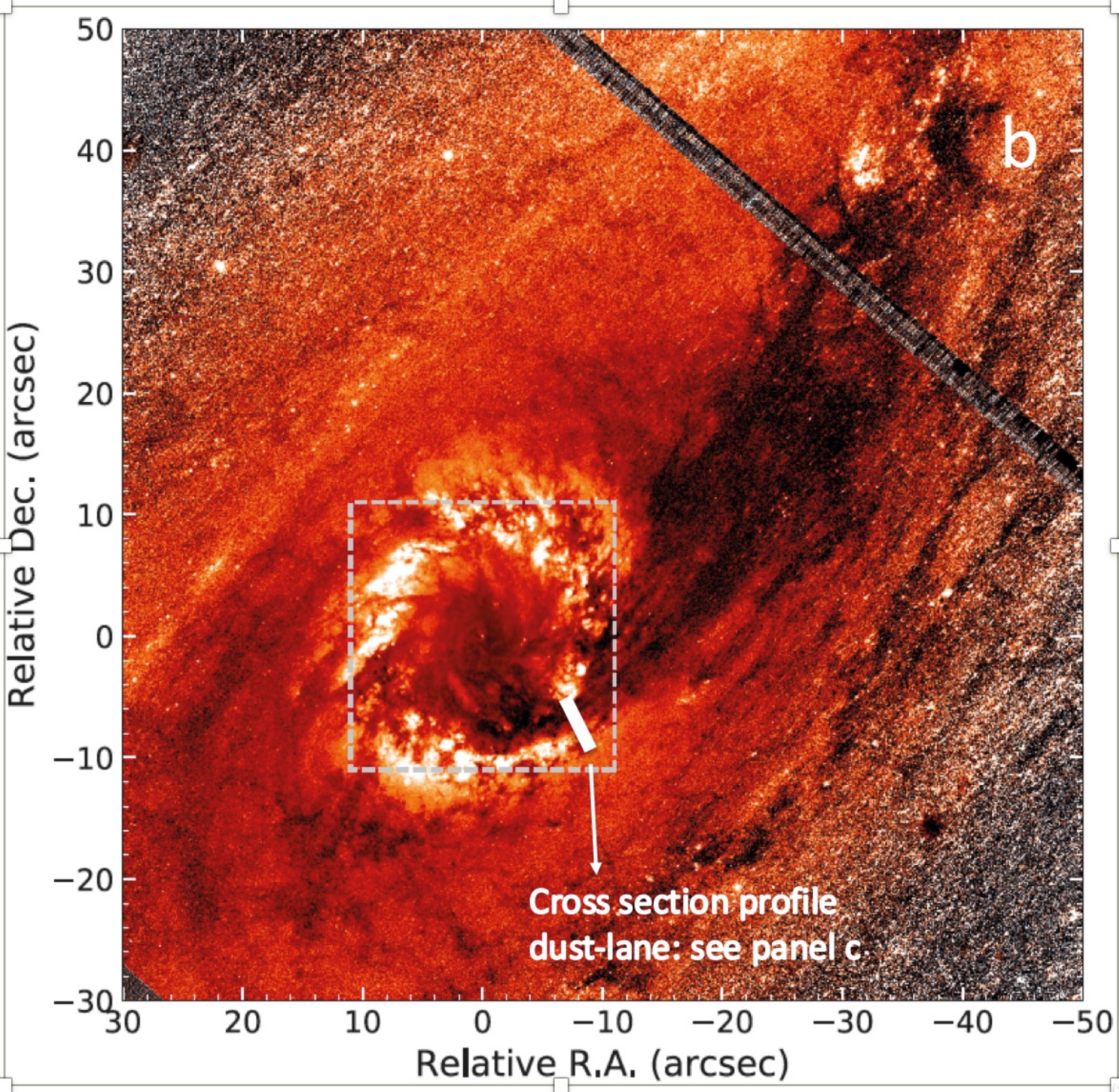} 
\includegraphics[width=0.5\textwidth]{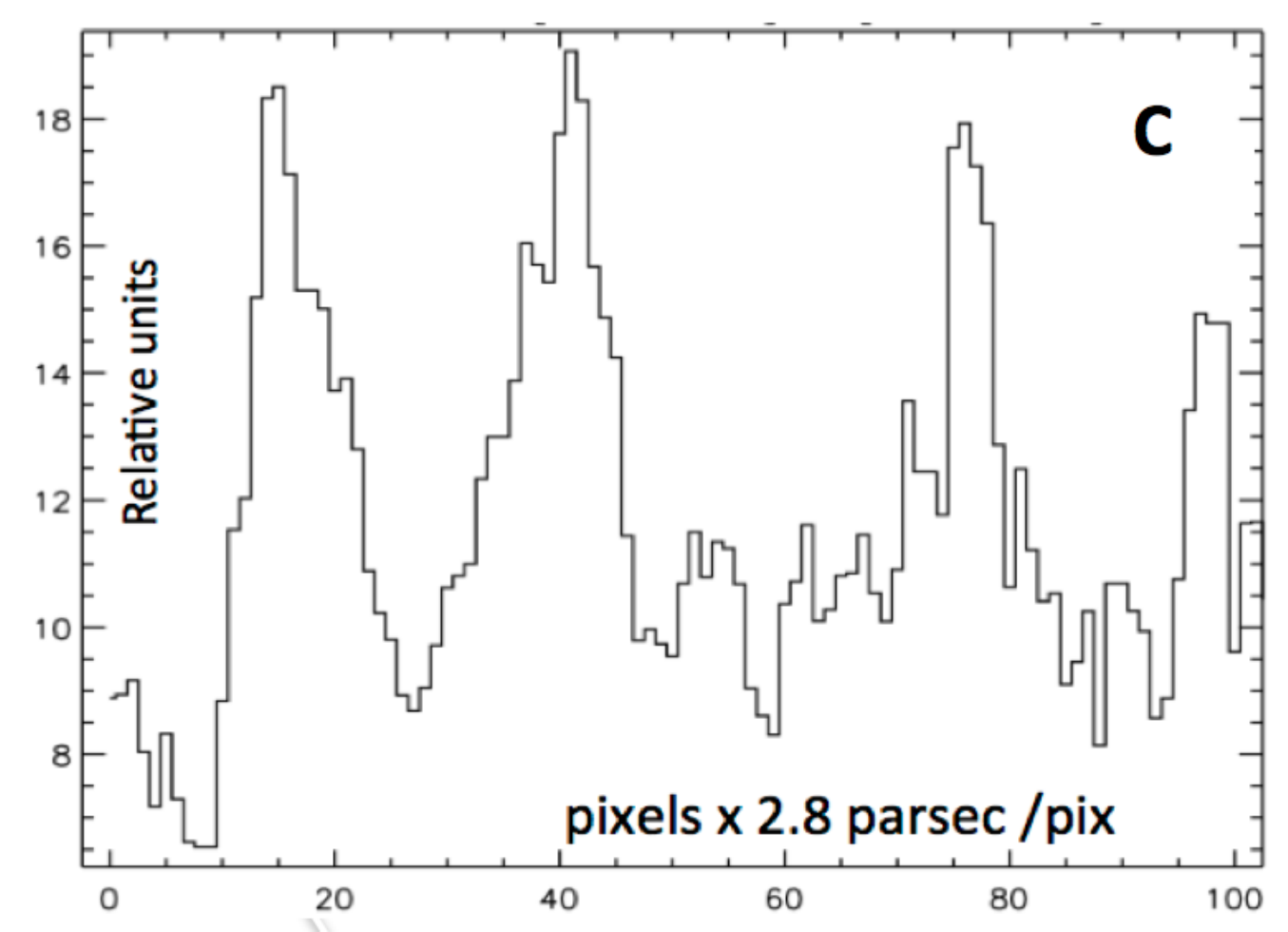}
\caption{From kpcs  distance to the centre of NGC 1097. a:  VLT/ VIMOS colour composite image of NGC 1097 (credit ESO), the white square marks the region used in (b); b:   Large-scale HST colour  map F814W / F438W used to produce the extinction map Av in Fig 1e, darker regions mark the dust location, the white square shows the field in Fig. 1e; c: a cross section through the dust-lane,  at the location indicated by the white line in panel b, that illustrates the filamentary  lane structure.
   North  up, East to the left in all panels.}\label{large}
\end{figure*}


\begin{thebibliography}{}

\bibitem[]{} Arsenault \& Roy 1988, A\&A, 201, 199
\bibitem[]{} Baldwin, J. et al. 1984, MNRAS, 210, 701
\bibitem[ ]{} Barvainis, R. 1987, ApJ 320, 537
\bibitem[ ]{} Barth, A.J., Ho, L.C., Filippenko, A.V., Sargent, W.L.  1995, AJ, 110, 1009
\bibitem[ ]{} Behrendt, M.; Burkert, A.; Schartmann, M. 2015, MNRAS 448, 1007
\bibitem[ ]{} Beck, R. et al. 2005, A\&A 444
\bibitem[ ]{} Bruzual, G. \& Charlot, S. 2003,  MNRAS 344, 1000
  
\bibitem[]{} Bruzual G., 2010, Philosophical Transactions RS London Series A, 368, 783
\bibitem[ ]{} Burkert, A. et al. 2010, ApJ 725, 2324
   \bibitem[ ]{} Burkert, A \& Hartmann, L., 2013, ApJ 773
  \bibitem[ ]{}  Burkert, A. 2017, Mem. S.A.It. Vol. 88, 533
   \bibitem[ ]{} Combes, F. et al. 2018, 	arXiv:1811.00984 
   \bibitem[ ]{} Combes, F. et al. 2014, A\&A 565, 97
\bibitem[ ]{} Davies, R. et al. 2009, ApJ 702
 \bibitem[ ]{} Dekel, A., Sari, R., \& Ceverino, D. 2009, ApJ, 703, 785

\bibitem[ ]{} Gillessen et al. 2012, Nat. 481,51
 \bibitem[]{} Espada et al. 2017, ApJ 843, 136
 \bibitem[]{}   Evans, N. J. et al. 2009, Ap. J. Suppl. 181, 321
 \bibitem[]{}Evans, N. J. 2017, AAS Meeting 230, id.307.01
\bibitem[ ]{} Fabian, A. 2012, ARAA 50, 455
\bibitem[ ]{} Fathi, K. et al. 2013, ApJL, 770, L27
 \bibitem[ ]{} Fern\'andez-Ontiveros et al. 2009, MNRAS L16
  \bibitem[]{}Fern\'andez-Ontiveros et al. 2012, Journal of Physics Conference Series, 372,012006 
\bibitem[]{}Friedli \& Benz 1993,  A\&A, 268, 65
\bibitem[]{}Gao \& Salomon 2004, Ap. J. 606, 271
\bibitem[]{} Gebhardt, K. et al. 2000, ApJ 539, L13
    \bibitem[]{}Gutkin J., Charlot S., Bruzual G., 2016, MNRAS, 462, 1757
        \bibitem[]{}Hollyhaed et al. 2015, MNRAS 449, 1106
    \bibitem[]{}Hsieh, P.-Y. et al. 2011, ApJ, 736, 129
 \bibitem[]{}Hummel et al. 1987, A\&A, 172, 32
    \bibitem[]{}Izumi et al. 2013, PASJ, 65, 1 L1600
    \bibitem[]{} Imanishi et al. 2016, ApJ 822L, 10
    \bibitem[]{}Kennicutt, R. \& Evans, N., 2012, ARA\&A, 50, 531
   \bibitem[]{}Koljonen et al. 2015ApJ 814,139
    \bibitem[]{}Kroupa P., 2001, MNRAS, 322, 231
 \bibitem[]{}Kulkarni, S. R.; Heiles, C. 1988, Galactic and extragalactic radio astronomy (2nd edition) (A89-40409 17-90). Berlin and New York, Springer-Verlag, p. 95-153
  \bibitem[]{}Lewis, K. T. \& Eracleous, M. 2006, ApJ, 642, 711
   \bibitem[]{}Li et al. 2015, ApJ 806, 150L
  \bibitem[]{}Malkan, M., Gorgian, \& Tam, R., 1998,  ApJS 117, 25
     \bibitem[]{}Martin et al. 2015, A\&A 573, A116
   \bibitem[]{}   Matsuda, T., \& Nelson, A. H. 1977, Nature, 266, 608
\bibitem[]{}Mason R. E. et al., 2007, ApJ, 659, 241
   \bibitem[]{} Mezcua et al. 2015, MNRAS 452, 4128
          \bibitem[]{}Mezcua \& Prieto, 2014, ApJ, 787, 62
           \bibitem[]{}    Mezcua et al. 2016, MNRAS 457L, 94
         \bibitem[]{}Mueller-Sanchez et al. 2009, ApJ 69L, 749
   \bibitem[]{}Ondrechen, M.P., van der Hulst \& Hummel, E., 1989, AJ, 342, 39
       \bibitem[]{} Osterbrock, D. 1989, Astrophysics of Gaseosus Nebulae and Active Galactic Nrouclei, Milli Valley, CA, Unv. Science Books
   \bibitem[]{}   Papaderos, P., Gomes,  J.M., Vilchez, J.M. et al. 2013, A\&A 551, L1
   \bibitem[]{}Prieto, M. A., Maciejewski, W., \& Reunanen, J. 2005, AJ, 130, 1472
   \bibitem[]{}Prieto et al. 2010, MNRAS, 402, 724
   \bibitem[]{}Prieto et al. 2014, MNRAS, 442, 2145
\bibitem[]{} Portegies Zwart S. F., McMillan, S. L.W., \& Gieles, M. 2010, ARA\&A, 48, 431
\bibitem[]{}Reunanen, J. Prieto, M.A. \& Siebenmorgen, R. 2010, MNRAS 402, 879
\bibitem[]{}Sakamoto, K.,  Okumura, S., Ishizuki, S., Scoville, N., 1999, ApJ 525, 691
\bibitem[]{}Savage \$ Mathis 1979, ARAA 17,33
      \bibitem[]{}     Sheth, K   et al. 2005, ApJ 632, 217
            \bibitem[]{}  Seale et al. 2012, ApJ 751, 42
    \bibitem[]{}     Sikora, M., Stawarz, ?.; Lasota, J.-P., 2007, ApJ 658, 815
            \bibitem[]{}   Silk,J., Rees, M.J. 1998, A\& 331, L1
 \bibitem[]{}Springel, W., di Matteo, T. \& Hernquist, L. 2005, MNRAS, 776
 \bibitem[]{}Storchi-Bergmann et al. 1997, ApJ, 489
   \bibitem[]{}Storchi-Bergmann et al. 2010, MNRAS 402, 819
 \bibitem[]{} Shlosman, Begelman and Frank 1990, Nat 345, 679
\bibitem[]{}  Tabatabei et al. 2017, Nat As 2, 83
\bibitem[]{} Tacconi et al. 2018,  ApJ 853, 179
\bibitem[]{}  Toomre  A. 1964, ApJ, 139, 1217
\bibitem[]{}Tully, R.B. 1988, Nearby Galaxies Catalog (Cambridge University Press)
\bibitem[]{}Vitt et al. 1992, ApJ, 393, 611
\bibitem[]{}Wofford A., et al., 2016, MNRAS, 457, 4296

\end{thebibliography}
 \end{document}